\documentclass[pdfa,conference,romanappendices]{IEEEtran}

\usepackage{multicol}
\usepackage{cite}
\usepackage{amsmath,amssymb,amsfonts}
\usepackage{graphicx}
\usepackage{textcomp}
\usepackage[font=small,skip=1pt]{caption}
\usepackage[caption=false,font=footnotesize]{subfig}
\usepackage{enumitem}
\usepackage[export]{adjustbox}
\usepackage{multirow}
\usepackage{float}
\usepackage{bigdelim}
\usepackage{makecell}
\usepackage{listings}
\usepackage[plain]{algorithm}
\usepackage[noend]{algpseudocode}
\usepackage{url}

\usepackage[breaklinks,colorlinks=true]{hyperref}
\usepackage{tlatex}
\usepackage{stmaryrd}
\usepackage{centernot}
\usepackage{comment}
\usepackage{empheq}
\usepackage{empheq}
\usepackage{amssymb}
\usepackage{pifont}
\usepackage[normalem]{ulem}
\usepackage{setspace}
\usepackage{scalerel}
\usepackage{bm}
\usepackage{xspace}
\usepackage{soul}

\usepackage{blindtext}

\usepackage{tikz}
\usetikzlibrary{shapes.geometric, arrows}
\usepackage{mathrsfs}
\usepackage{relsize}

\makeatletter
\newcommand{\pushright}[1]{\ifmeasuring@#1\else\omit\hfill$\displaystyle#1$\fi\ignorespaces}
\newcommand{\pushleft}[1]{\ifmeasuring@#1\else\omit$\displaystyle#1$\hfill\fi\ignorespaces}
\makeatother

\renewenvironment{comment}{}{}

\newcommand{\svalue}{\fin{value}}
\newcommand{\sacceptor}{\fin{acceptor}}
\newcommand{\squorum}{\fin{quorum}}
\newcommand{\sballot}{\fin{ballot}}
\newcommand{\sinstance}{\fin{instances}}

\makeatletter
\def\@citecolor{blue}%
\def\@urlcolor{blue}%

\def\orcidID#1{\smash{\href{http://orcid.org/#1}{\protect\raisebox{-1.25pt}{\protect\includegraphics{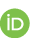}}}}}
\makeatother

\usepackage{bbding}

\newcommand{\tlasize}{\fontsize{8pt}{10pt}\selectfont}
\newcommand{\eqsize}{}

\newcommand{\asize}{\small}

\definecolor{nonecolor}{HTML}{666666}
\definecolor{highcolor}{HTML}{790C0C}
\definecolor{failcolor}{HTML}{666666}

\definecolor{boxshade}{HTML}{DCDCDC}

\definecolor{hlcolor}{HTML}{DCDCDC}

\definecolor{commentcolor}{HTML}{787c82}

\usepackage{mdframed}
\newmdenv[
backgroundcolor=hlcolor,
topline=false,
bottomline=false,
leftline=false,
rightline=false,
]{shaded}

\definecolor{color1}{HTML}{E8f0f9}
\definecolor{color2}{HTML}{FBEAEA}

\newmdenv[
leftmargin=-.15cm,
rightmargin=-.15cm,
skipabove=0cm,
skipbelow=0cm,
innerleftmargin=0.15cm,
innerrightmargin=0cm,
innertopmargin=0.15cm,
innerbottommargin=0.15cm,
backgroundcolor=color1,
topline=false,
bottomline=false,
leftline=false,
rightline=false,
]{shaded1}

\newmdenv[
backgroundcolor=color1,
topline=false,
bottomline=false,
leftline=false,
rightline=false,
]{shaded2}

\newmdenv[
leftmargin=-.15cm,
rightmargin=-.15cm,
skipabove=0cm,
skipbelow=0cm,
innerleftmargin=0.15cm,
innerrightmargin=0cm,
innertopmargin=0.15cm,
innerbottommargin=0.15cm,
backgroundcolor=color2,
topline=false,
bottomline=false,
leftline=false,
rightline=false,
]{shaded3}

\newmdenv[
backgroundcolor=color2,
topline=false,
bottomline=false,
leftline=false,
rightline=false,
]{shaded4}

\algnewcommand\algorithmicforeach{\textbf{for each}}
\algdef{S}[FOR]{ForEach}[1]{\algorithmicforeach\ #1\ \algorithmicdo}

\newcommand{\fin}[1]{{\tt #1}}

\newcommand{\blind}[1]{}
\newcommand{\hide}[1]{}

\usepackage{mathtools}

\algnewcommand{\algorithmicgoto}{\textbf{go to} Line}%
\algnewcommand{\Goto}[1]{\algorithmicgoto~\ref{#1}}%

\algnewcommand{\algorithmicnot}{\textbf{not}}%
\algnewcommand{\Not}{\algorithmicnot}%

\newcommand{\icpo}{IC3PO\xspace}
\newcommand{\updr}{UPDR\xspace}
\newcommand{\folic}{fol-ic3\xspace}
\newcommand{\ifour}{I4\xspace}
\newcommand{\swiss}{SWISS\xspace}
\newcommand{\distai}{DistAI\xspace}

\newcommand{\voting}{\textit{Voting}\xspace}
\newcommand{\paxos}{\textit{Paxos}\xspace}
\newcommand{\multipaxos}{\textit{MultiPaxos}\xspace}
\newcommand{\flexiblepaxos}{\textit{FlexiblePaxos}\xspace}
\newcommand{\paxossimple}{\textit{SimplePaxos}\xspace}
\newcommand{\paxosabstract}{\textit{ImplicitPaxos}\xspace}

\newcommand{\ev}{V!}
\newcommand{\eps}{S!}
\newcommand{\epa}{I!}
\newcommand{\ep}{P!}
\newcommand{\emp}{M!}

\newcommand{\eh}{H!}
\newcommand{\el}{L!}

\newcommand{\init}{Init}
\newcommand{\nex}{Next}
\newcommand{\prop}{Safety}

\usepackage{scalerel,stackengine}
\stackMath
\newcommand\reallywidehat[1]{%
\savestack{\tmpbox}{\stretchto{%
  \scaleto{%
    \scalerel*[\widthof{\ensuremath{#1}}]{\kern-.6pt\bigwedge\kern-.6pt}%
    {\rule[-\textheight/2]{1ex}{\textheight}}
  }{0.3\textheight}%
}{0.4ex}}%
\stackon[1pt]{#1}{\tmpbox}%
}
\newcommand{\fnex}{\reallywidehat{\nex}}

\newcommand{\s}{\fin{s}}

\newcommand{\epr}{$\mathtt{EPR}$\xspace}
\newcommand{\orig}{$\mathtt{ORIGINAL}$\xspace}

\begin{document}

\title{Towards an Automatic Proof of Lamport's Paxos}

\author{\IEEEauthorblockN{Aman Goel \orcidID{0000-0003-0520-8890}}
\IEEEauthorblockA{
\textit{University of Michigan, Ann Arbor}\\
amangoel@umich.edu}
\and
\IEEEauthorblockN{Karem A. Sakallah \orcidID{0000-0002-5819-9089}}
\IEEEauthorblockA{
\textit{University of Michigan, Ann Arbor}\\
karem@umich.edu}
}
\maketitle
\thispagestyle{plain}
\pagestyle{plain}

\begin{abstract}
Lamport's celebrated Paxos consensus protocol is generally viewed as a complex hard-to-understand algorithm. Notwithstanding its complexity, in this paper, we take a step towards automatically proving the safety of Paxos by taking advantage of three structural features in its specification: \textit{spatial regularity} in its unordered domains, \textit{temporal regularity} in its totally-ordered domain, and its \textit{hierarchical composition}. By carefully integrating these structural features in \icpo, a novel model checking algorithm, we were able to infer an inductive invariant that identically matches the human-written one previously derived with significant manual effort using interactive theorem proving. While various attempts have been made to verify different versions of Paxos, to the best of our knowledge, this is the first demonstration of an automatically-inferred inductive invariant for Lamport's original Paxos specification. We note that these structural features are not specific to Paxos and that \icpo can serve as an automatic general-purpose protocol verification tool.
\end{abstract}


\section{Introduction}
\label{sec:intro}
\noindent In this paper, we focus on proving the \textit{safety} of distributed protocols like Paxos~\cite{lamport1998part,lamport2001paxos} which form the basis for implementing many efficient and highly fault-tolerant distributed services~\cite{burrows2006chubby,paxos_alive,autopilot}.
Developed by Lamport, the Paxos consensus protocol allows a set of processes to communicate with each other by exchanging messages and reach agreement on a single value.
Verifying the correctness of such a concurrent system requires the derivation of a \textit{quantified inductive invariant} that, together with the protocol specification, acts as an inductive proof of its safety under all possible system behaviors.

Several manual or semi-automatic verification techniques based on interactive theorem proving~\cite{de2000revisiting,chaudhuri2010tla,tla_proofs,nipkow2002isabelle} have been proposed to derive a safety proof for Paxos.
Chand et al.~\cite{chand2016formal} formally verified the TLA+~\cite{lamport2002specifying} specification of Paxos by manually deriving a proof using the TLAPS proof assistant~\cite{chaudhuri2010tla}.
Padon et al.~\cite{padon2017paxos} used the Ivy~\cite{padon2016ivy} verifier, which requires a user to manually refine automatically-generated counterexamples-to-induction, to obtain an inductive invariant for a simplified version of Paxos in the decidable EPR fragment~\cite{piskac2010deciding} of first-order logic. 
The approaches in~\cite{lamport2011byzantizing,hawblitzel2015ironfleet,Wilcox2015Verdi,merz2019formal,kragl2020refinement} are examples of manually-derived \textit{refinement proofs}~\cite{abadi1991existence,lamport1994temporal,lamport1996refinement,garland2000using} that show how a low-level implementation refines a high-level specification.
All these methods, however, require a detailed understanding of the intricate inner workings of the protocol and entail significant manual effort to guide proof development.

\begin{figure}[!tb]
\captionsetup{type=figure}
\centering
        \includegraphics[width=\linewidth]{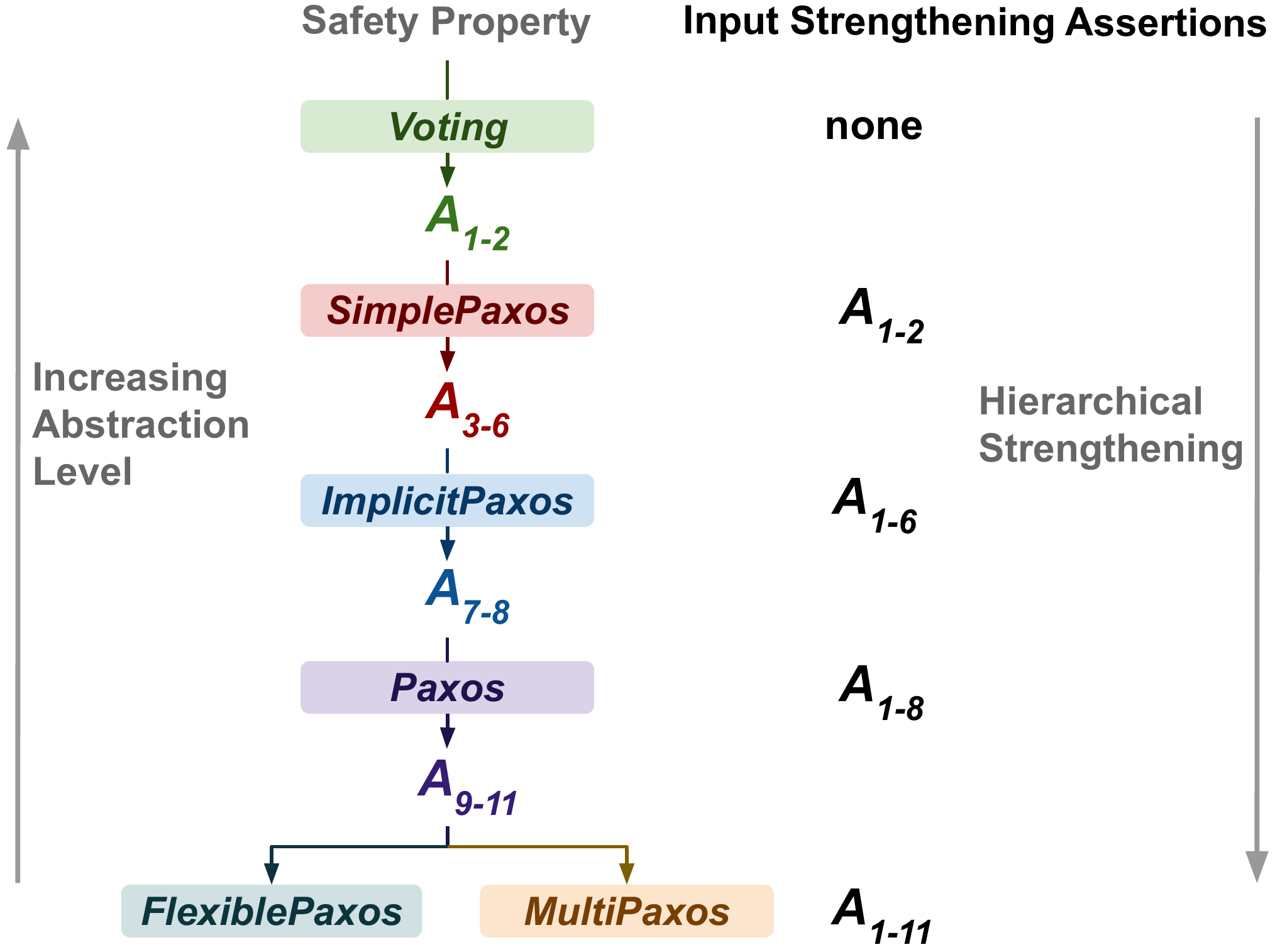}
\caption{Hierarchical strengthening of Paxos and its variants. Each level uses all strengthening assertions above that level as input, and outputs the required remaining assertions, altogether inferring the inductive invariant at each level.}
\label{fig:paxos_refine}
\end{figure}

In contrast, we propose an approach, implemented in the IC3PO protocol verifier, to \textit{automatically infer the required inductive invariant} for an unbounded distributed protocol by adding three simple extensions to the finite-domain IC3/PDR~\cite{bradley2011sat,een2011efficient} incremental induction algorithm for model checking~\cite{clarke2009model}.
\textit{Symmetry boosting}, introduced in~\cite{goel2021on}, takes advantage of a protocol's \textit{spatial} regularity to automatically infer quantified strengthening assertions that reflect the protocol's structural symmetries. This paper describes \textit{range boosting} and \textit{hierarchical strengthening} which take advantage, respectively, of a protocol's \textit{temporal} regularity and hierarchical structure, and demonstrates how IC3PO was used to automatically obtain an inductive invariant for Paxos using the four-level hierarchy shown in Figure~\ref{fig:paxos_refine}.

\noindent Our main contributions are:
\begin{itemize}
    \item[--] A \textit{range boosting} technique that extends incremental induction to utilize the \textit{temporal regularity} in totally-ordered domains, and thus, enables automatic invariant inference for protocols with even \textit{infinite-state} processes.
    
    \item[--] A \textit{hierarchical strengthening} approach to derive the required inductive invariant in a top-down step-wise procedure for hierarchically-specified distributed protocols through incremental induction extended with symmetry and range boosting, by automatically verifying high-level abstractions first and using invariants of these higher-level abstractions as \textit{strengthening assertions} to derive the inductive invariant for the detailed lower-level protocol.
    
    \item[--] Safety verification of \textit{Lamport's Paxos algorithm}, both single- and multi-decree Paxos, through the derivation of a compact, human-readable inductive proof that is automatically inferred using \icpo, resulting in a drastic reduction in verification effort compared to previous approaches~\cite{paxos_proof,hawblitzel2015ironfleet,taube2018modularity}.
\end{itemize}
\noindent The paper is structured as follows: \S\ref{sec:prelim} presents preliminaries. \S\ref{sec:range} and \S\ref{sec:hierarchical} describe range boosting and hierarchical strengthening. \S\ref{sec:paxos} details the four-level hierarchy we used to prove Paxos and \S\ref{sec:paxos_hr} is a record of the IC3PO run showing the actual assertions it inferred at each level of the hierarchy. \S\ref{sec:paxos_discuss} discusses some of the features and interesting details on this automatically-generated proof. Experimental comparisons with other approaches are provided in \S\ref{sec:exp} and the paper  concludes with a brief survey of related work in \S\ref{sec:related} and a discussion of future directions in \S\ref{sec:conclude}.

\section{Preliminaries}
\label{sec:prelim}

\subsection{Notation}
\label{subsec:notation}
\noindent
We will use $\init, \nex$, and $\prop$ to denote the  quantified formulas that specify, respectively, a protocol's initial states, its transition relation, and the safety property that is required to hold on all reachable states.
We use primes (e.g., $\varphi'$) to represent a formula after a single transition step. The notation $\ev A$ (resp. $\eps A$, $\epa A$, and $\ep A$) means that assertion $A$ was inferred by IC3PO for the \voting (resp. \paxossimple, \paxosabstract, and \paxos) protocol.

As an example, consider a protocol $\mathcal{P}$ with two  sorts, a symmetric sort $\fin{aSort}$ and a totally-ordered sort  $\fin{bSort}$, along with relations $p(\fin{aSort}, \fin{bSort})$ and $q(\fin{bSort})$ defined on these sorts.
Viewed as a parameterized system $\mathcal{P}(\fin{aSort}, \fin{bSort})$, we can specify its finite instance $\mathcal{P}(3, 4)$ as:
{
\eqsize
\begin{align}
\mathcal{P}(3, 4) : \hspace{20pt} &\fin{aSort_3} \triangleq \{ \fin{a_1}, \fin{a_2}, \fin{a_3} \}  \nonumber\\
&\fin{bSort_4} \triangleq [ \fin{b_{min}, b_1, b_2, b_{max}} ] \label{eq:ex_size}
\end{align}
}%
where $\fin{aSort_3}$ represents the finite symmetric sort of this instance defined as a set of arbitrarily-named distinct constants, while the finite totally-ordered sort $\fin{bSort_4}$ is composed of a list of ordered constants, i.e., $\fin{b_{min} < b_1 < b_2 < b_{max}}$.
This instance can be encoded using twelve $p$ and four $q$ $\BOOLEAN$ state variables. A \textit{state} of this instance corresponds to a complete assignment to these $16$ state variables, with a total state-space size of $2^{16}$.
We will use $\fnex$ instead of $\nex$ to denote the transition relation of the finite instance.

\subsection{Clause Boosting and Quantifier Inference}
\label{subsec:ic3po}
\noindent The basic framework for inferring the quantified assertions required to prove protocol safety is described in~\cite{goel2021on}. It extends the finite IC3/PDR incremental induction algorithm by \textit{boosting} its clause learning during the 1-step backward reachability checks performed through Satisfiability Modulo Theories (SMT)~\cite{BarFT-SMTLIB} solving. Specifically, a clause $\varphi$ is learned in (and refines) frame $F_i$ if the 1-step query $\psi_i := F_{i-1} \wedge \fnex \wedge [\neg \varphi']$ is unsatisfiable. This means that cube $\neg \varphi$ in frame $F_i$ is unreachable from frame $F_{i-1}$.
Boosting refers to: a) ``growing'' $\varphi$ to a set of clauses that also satisfy this \textit{unreachability constraint} from frame $F_{i-1}$, and b) refining the frame $F_i$ with the entire clause set instead of just $\varphi$.
Such boosting accelerates the convergence of incremental induction but, more importantly, makes it possible, under some regularity assumptions, to represent this set of clauses by a \textit{single logically-equivalent quantified clause} $\Phi$ and is the key to generalizing the results of such finite analysis to unbounded domains.

\subsection{Symmetric Boosting and Quantifier Inference}
\noindent
 Protocols that are strictly specified in terms of symmetric sorts can be characterized as having \textit{spatial} regularity. For example, the constants in a sort representing a finite set of $k$ identical processes are essentially indistinguishable \textit{replicas} that can be permuted arbitrarily without changing the protocol behavior. A learned clause $\varphi$ parameterized by the constants of such a sort can be boosted by permuting its constants in all possible $k!$ ways yielding a set of symmetrically-equivalent clauses, i.e., its symmetry \textit{orbit} $\varphi^{Sym_k}$ under the full symmetric group $Sym_k$. By construction, all clauses in $\varphi$'s orbit automatically satisfy the unreachability constraint without the need to perform additional 1-step queries.
 Furthermore, the quantified clause $\Phi$ that encodes $\varphi$'s orbit is algorithmically constructed by a syntactic analysis of $\varphi$'s structure, and can involve complex universal and existential quantifier alternations over both state and non-state (auxiliary) variables. The reader is referred to~\cite{goel2021on,goel2021on_arxiv} for the complete details of the connection between symmetry and quantification and the procedure for quantifier inference.
 
\subsection{Finite Convergence}
\noindent
When a boosted finite incremental induction run terminates, it either produces a finite counterexample demonstrating that the specified safety property fails, or produces a set of quantified assertions $A_1, \cdots, A_n$ that yield the inductive invariant $inv = Safety \wedge A_1 \wedge \cdots \wedge A_n$ proving safety for the given finite size. At this point, an algorithmic \textit{finite convergence} procedure is invoked to check if the current instance size has captured all possible protocol behaviors and, if not, to systematically increase the finite instance size until protocol behavior saturates and the cutoff size is reached~\cite{pnueli2001automatic,arons2001parameterized, zuck2004model,balaban2005iiv,namjoshi2007symmetry}.

\section{Range Boosting}
\label{sec:range}
\noindent
Clause boosting is not limited to clauses that are parameterized by the constants of symmetric sorts, and can be extended to clauses whose literals depend on the constants of totally-ordered sorts such as  ballot, round, epoch, etc., that are used to model the temporal order of events in a distributed protocol. However, the boosting procedure for such clauses differs from symmetric boosting in two ways: a) the ordering relation between totally-ordered constants must be explicitly preserved, and b) adherence of a boosted clause to the unreachability constraint is not guaranteed and must be explicitly checked with a 1-step backward reachability query.

We extended \icpo with a \textit{range boosting} procedure that complements its symmetry boosting mechanism, allowing it to transparently handle protocols with both symmetric and totally-ordered sorts.

Let  $\varphi$ be a clause that is parameterized by totally-ordered constants and let $\varphi^{Ordered}$ denote those variants of $\varphi$ that are obtained by ordering-compliant permutations of its constants. 
Clause $\varphi$ is boosted by making 1-step backward reachability queries on $\varphi^{Ordered}$ to identify its \textit{safe} subset $\varphi^{Safe}$,  i.e., those variants that satisfy the unreachability constraint.

For example, consider the following clause $\varphi_1$ defined on the finite instance $\mathcal{P}(3,4)$ from (\ref{eq:ex_size}):
{
\eqsize
\begin{align}
\varphi_1 =&~ p(\fin{a_1, b_1}) \vee q(\fin{b_2}) \label{eq:clause1}
\end{align}
}%
Since $\varphi_1$ contains two ordered constants $(\fin{b_1}, \fin{b_2})$, it has six ordering-compliant variants $ \fin{(b_{min},b_1)},$ $\fin{(b_{min},b_2)},$ $\fin{(b_{min},b_{max})},$ $\fin{(b_1,b_2)},$ $\fin{(b_1,b_{max})}$, and  $\fin{(b_2,b_{max})}$. However only three of these variants end up satisfying the unreachability constraint yielding the following safe subset of $\varphi_1^{Ordered}$:

{
\eqsize
\begin{align}
\varphi_1^{Safe} =&~ [~ p(\fin{a_1, b_1}) \vee q(\fin{b_2}) \hspace{8pt} ~]~ \wedge \nonumber\\
&~ [~ p(\fin{a_1, b_1}) \vee q(\fin{b_{max}}) ~]~ \wedge \nonumber\\
&~ [~ p(\fin{a_1, b_2}) \vee q(\fin{b_{max}}) ~]~ \label{eq:clause1_orbit2}
\end{align}
}%

\noindent
The inferred quantified clause that encodes these three clauses is now constructed using two universally-quantified variables $X_1, X_2 \in \fin{bSort_4}$ that replace $\fin{b_1}$ and $\fin{b_2}$ in $\varphi_1$ and expressed as an implication whose antecedent specifies a constraint over the ordered ``range'' $\fin{b_{min}} < X_1 < X_2$ that must be satisfied by the quantified variables:
{
\eqsize
\begin{align}
& \Phi_1 =~ \forall X_1, X_2 \in \fin{bSort_4}: \nonumber\\
& \hspace{20pt} (\fin{b_{min}} < X_1) \wedge (X_1 < X_2) \to [~ p(\fin{a_1}, X_1) \vee q(X_2) ~] \label{eq:clause1_quant2}
\end{align}
}%

\noindent
In general, a clause that is parameterized by $k$ constants from a totally-ordered domain whose size is greater than $k$ can be range-boosted and encoded by a universally-quantified predicate with $k$ variables which is expressed as an implication whose antecedent is a range constraint that evaluates to true for just those combinations of the $k$ variables that correspond to safe variants of $\varphi$.

This procedure extends easily to the case of multiple totally-ordered domains as well, allowing range boosting to be performed independently for each such domain in \textit{any} order since constants from different domains do not interfere with each other.

\section{Hierarchical Strengthening}
\label{sec:hierarchical}
\noindent
As advocated in~\cite{lamport1995write}, hierarchical structuring is an effective way to manage complexity during manual proof development. It can also be easily incorporated in the \icpo style of invariant generation based on symmetry and range boosting.

Given a low-level specification $L$ that implements a high-level specification $H$, i.e., $L \prec H$, hierarchical strengthening starts by automatically deriving strengthening assertions $\eh A^H $ that, together with the safety property $\eh \prop$, proves the safety of $H$. It then maps and propagates $\eh A^H$ to $L$, denoted as $\el A^H$, and proceeds to prove the strengthened property $\el \prop \wedge \el A^H$ in $L$ by deriving any additional assertions $\el A^L$ needed to establish the safety of $L$. The underlying assumption in this procedure is that proving $H$ is much easier than proving $L$ directly, and that any assertions derived to prove $H$ are also applicable, with suitable mapping, to $L$. The final inductive invariant that proves $L$ will, thus, have the form $\el inv = (\el \prop \wedge \el A^H) \wedge \el A^L$ which can be interpreted as reducing the complexity of $L$'s proof by strengthening its safety property with assertions derived for $H$.

Such strengthening can be extended to a $k$-level hierarchy $H \prec M_1 \prec \cdots \prec M_{k-2} \prec L$, where $M_1$ to $M_{k-2}$ are suitably-defined intermediate levels between $H$ and $L$. 
This, in turn, allows single-level automatic verification techniques based on incremental induction, like \icpo, to scale to complex protocols like \paxos, by step-wise verifying higher-level abstractions first and using their auto-generated proofs to incrementally build the proof for the lower-level protocol.

\section{Hierarchical Specification of Paxos}
\label{sec:paxos}
\noindent
This section describes in detail the multi-level hierarchical structure of the Paxos protocol, as shown earlier in Figure~\ref{fig:paxos_refine}.

\begin{algorithm}[!tb]
\batchmode 
\tlatex
\setstretch{0.93}
\tlasize
\setboolean{shading}{true}
\@x{\makebox[5pt][r]{\scriptsize \hspace{1em}}}\moduleLeftDash\@xx{ {\MODULE} \text{\voting}}\moduleRightDash\@xx{\makebox[5pt][r]{\scriptsize \hspace{1em}}}%

\@pvspace{4.0pt}%
\@x{\makebox[12pt][r]{\scriptsize 1\hspace{0.8em}} {\CONSTANTS} \svalue , \sacceptor , \squorum}%

\@pvspace{4.0pt}%
\@x{\makebox[12pt][r]{\scriptsize 2\hspace{0.8em}} \sballot \.{\defeq} Nat
 \.{\cup} \{ -1 \}}%

\@pvspace{4.0pt}%
\@x{\makebox[12pt][r]{\scriptsize 3\hspace{0.8em}} {\VARIABLES} votes , maxBal}%
\@pvspace{4.0pt}%
\@x{\makebox[12pt][r]{\scriptsize 4\hspace{0.8em}} votes\@s{9} \.{\in} ( \sacceptor \times \sballot \times \svalue) \.{\rightarrow}
 {\BOOLEAN}}%
\@pvspace{4.0pt}%
\@x{\makebox[12pt][r]{\scriptsize \hspace{0.8em}} maxBal \.{\in} \sacceptor \.{\rightarrow}
 \sballot}%

\@pvspace{4.0pt}%
\@x{\makebox[12pt][r]{\scriptsize 5\hspace{0.8em}} {\ASSUME}\@s{4} \.{\land}  \@s{4} \A\, Q \.{\in} \squorum \.{:} Q \subseteq \sacceptor } %
\@x{\makebox[12pt][r]{\scriptsize \hspace{0.8em}} \@s{34} \.{\land}  \@s{4} \A\, Q_1 , Q_2 \.{\in} \squorum \.{:} Q_1 \.{\cap} Q_2 \.{\neq} \{ \} } %

\@pvspace{4.0pt}%
\@x{\makebox[12pt][r]{\scriptsize 6\hspace{0.8em}} chosenAt(b, v) \.{\defeq} \E\, Q \.{\in} \squorum \.{:} \A\, A \.{\in} Q \.{:} votes(
 A , b , v )}
 
\@pvspace{4.0pt}%
\@x{\makebox[12pt][r]{\scriptsize 7\hspace{0.8em}} chosen(v) \.{\defeq} \E\, B \.{\in} \sballot \.{:} chosenAt(B , v )}
 
\@pvspace{4.0pt}%
\@x{\makebox[12pt][r]{\scriptsize 8\hspace{0.8em}} showsSafeAt ( q , b , v ) \.{\defeq}}%
\@x{\@s{25} \.{\land} \@s{4} \A\, A \.{\in} q \.{:} maxBal(A) \.{\geq} b}%
\@x{\@s{25} \.{\land} \@s{4} \E\, C \.{\in} \sballot \.{:}}%
\@x{\@s{40} \.{\land} \@s{4} ( C < b)}%
\@x{\@s{40} \.{\land} \@s{4} ( C \.{\neq} -1 ) \.{\rightarrow} \E\, A \.{\in} q \.{:} votes( A , C , v )}%
\@x{\@s{40} \.{\land} \@s{4} \A\, D \.{\in} \sballot \.{:}}%
\@x{\@s{65} (C < D < b) \.{\rightarrow}}%
 \@x{\@s{85} \A\, A \.{\in} Q \.{:} \A\, V \.{\in} \svalue \.{:} \neg votes( A , D , V )}%
 
\@pvspace{4.0pt}%
\@x{\makebox[12pt][r]{\scriptsize 9\hspace{0.8em}} isSafeAt(b , v) \.{\defeq} \E\, Q \.{\in} \squorum \.{:} showsSafeAt(Q , b , v )}

\@pvspace{4.0pt}%
\@x{\makebox[12pt][r]{\scriptsize 10\hspace{0.8em}} IncreaseMaxBal ( a , b ) \@s{2} \.{\defeq}}
\@x{\@s{25} \.{\land} \@s{4} b \.{\neq} -1 \@s{4} \.{\land} \@s{4} b > maxBal(a)}
\@x{\@s{25} \.{\land} \@s{4} maxBal \.{'} \.{=} [ maxBal {\EXCEPT} {\bang} [ a
 ] \.{=} b ]}%
\@x{\@s{25} \.{\land} \@s{4} \.{\UNCHANGED} votes}%

\@pvspace{4.0pt}%
\@x{\makebox[12pt][r]{\scriptsize 11\hspace{0.8em}} VoteFor ( a , b , v ) \@s{2} \.{\defeq}}
\@x{\@s{25} \.{\land} \@s{4} b \.{\neq} -1 \@s{4} \.{\land} \@s{4} maxBal(a) \leq b}
\@x{\@s{25} \.{\land} \@s{4} \A\, V \.{\in} \svalue \.{:} \neg votes( a , b , V )}
\@x{\@s{25} \.{\land} \@s{4} \A\, C \.{\in} \sacceptor \.{:} }
\@x{\@s{65} (C \.{\neq} a) \rightarrow }
\@x{\@s{85} \A\, V \.{\in} \svalue \.{:} votes(C, b, V) \rightarrow (V = v)}
\@x{\@s{25} \.{\land} \@s{4} isSafeAt( b , v )}
\@x{\@s{25} \.{\land} \@s{4} votes \.{'} \@s{9} \.{=} [ votes {\EXCEPT} {\bang} [a, b, v] \.{=} \top ]}%
\@x{\@s{25} \.{\land} \@s{4} maxBal \.{'} \.{=} [ maxBal {\EXCEPT} {\bang} [ a ] \.{=} b ]}%

\@pvspace{4.0pt}%
\@x{\makebox[12pt][r]{\scriptsize 12\hspace{0.8em}} \init \@s{0} \.{\defeq}}
\@x{\@s{15} \.{\land} \@s{2} \.\A\, A \.{\in} \sacceptor \.{:} B \.{\in} \sballot \.{:} V \.{\in} \svalue \.{:} {\neg} votes ( A , B , V )} %
\@x{\@s{15} \.{\land} \@s{2} \.\A\, A \.{\in} \sacceptor \.{:} maxBal(A) = -1} %

\@pvspace{4.0pt}%
\@x{\makebox[12pt][r]{\scriptsize 13\hspace{0.8em}} \nex \@s{0} \.{\defeq} \@s{0} \.\E\, A {\in} \sacceptor , B {\in} \sballot, V {\in} \svalue \,{:}\, }
\@x{\@s{55} IncreaseMaxBal(A , B) \.{\lor} VoteFor(A , B , V)}%

\@pvspace{4.0pt}%
 \@x{\makebox[12pt][r]{\scriptsize 14\hspace{0.8em}} \resizebox{0.95\linewidth}{!}{$\prop \@s{0} \.{\defeq} \@s{0} \.\A\, V_1, V_2 {\in} \svalue \.{:} chosen(V_1) \.{\land} chosen(V_2) \.{\rightarrow} V_1 \.{=} V_2$}}%

 \@pvspace{4.0pt}%
\@x{\makebox[5pt][r]{\scriptsize \hspace{1em}}}\bottombar\@xx{\makebox[5pt][r]{\scriptsize \hspace{1em}}}%
\captionof{figure}{Lamport's \voting protocol in pretty-printed TLA+}
\label{fig:voting}
\end{algorithm}%

\begin{algorithm}[!tb]
\batchmode 
\tlatex
\setstretch{0.93}
\tlasize
\setboolean{shading}{true}
\@x{\makebox[5pt][r]{\scriptsize \hspace{1em}}}\moduleLeftDash\@xx{ {\MODULE} \text{\paxos}}\moduleRightDash\@xx{\makebox[5pt][r]{\scriptsize \hspace{1em}}}%

\@pvspace{4.0pt}%
\@x{\makebox[10pt][r]{\scriptsize 1\hspace{0.8em}} {\CONSTANTS} \svalue , \sacceptor , \squorum}%

\@pvspace{4.0pt}%
\@x{\makebox[10pt][r]{\scriptsize 2\hspace{0.8em}} \sballot \.{\defeq} Nat
 \.{\cup} \{ -1 \}}%

\@pvspace{4.0pt}%
\@x{\makebox[10pt][r]{\scriptsize 3\hspace{0.8em}} {\VARIABLES} msg1a , msg1b , msg2a , msg2b , maxBal}%
\@x{\@s{50} maxVBal , maxVal}%

\@pvspace{4.0pt}%
\@x{\makebox[10pt][r]{\scriptsize 4\hspace{0.8em}} msg1a \@s{12} \.{\in} \sballot \.{\rightarrow}
 {\BOOLEAN}}%

\@pvspace{4.0pt}%
\@x{\makebox[10pt][r]{\scriptsize \hspace{0.8em}} msg1b \@s{12} \.{\in} (\sacceptor \times \sballot \times \sballot \times \svalue) \.{\rightarrow}
 {\BOOLEAN}}%

\@pvspace{4.0pt}%
\@x{\makebox[10pt][r]{\scriptsize \hspace{0.8em}} msg2a \@s{12} \.{\in} (\sballot \times \svalue) \.{\rightarrow}
 {\BOOLEAN}}%

\@pvspace{4.0pt}%
\@x{\makebox[10pt][r]{\scriptsize \hspace{0.8em}} msg2b \@s{12} \.{\in} (\sacceptor \times \sballot \times \svalue) \.{\rightarrow}
 {\BOOLEAN}}%
 
\@pvspace{4.0pt}%
\@x{\makebox[10pt][r]{\scriptsize \hspace{0.8em}} maxBal\@s{7} \.{\in} \sacceptor \.{\rightarrow}
 \sballot}%

\@pvspace{4.0pt}%
\@x{\makebox[10pt][r]{\scriptsize \hspace{0.8em}} maxVBal \.{\in} \sacceptor \.{\rightarrow} \sballot}%

\@pvspace{4.0pt}%
\@x{\makebox[10pt][r]{\scriptsize \hspace{0.8em}} maxVal\@s{7} \.{\in} \sacceptor \.{\rightarrow}
 \svalue}%

\@pvspace{4.0pt}%
\@x{\makebox[10pt][r]{\scriptsize \hspace{0.8em}} none\@s{17} \.{\in} \svalue}%

\@pvspace{4.0pt}%
\@x{\makebox[10pt][r]{\scriptsize 5\hspace{0.8em}} {\ASSUME}\@s{4} \.{\land}  \@s{4} \A\, Q \.{\in} \squorum \.{:} Q \subseteq \sacceptor } %
\@x{\makebox[10pt][r]{\scriptsize \hspace{0.8em}} \@s{34} \.{\land}  \@s{4} \A\, Q_1 , Q_2 \.{\in} \squorum \.{:} Q_1 \.{\cap} Q_2 \.{\neq} \{ \} } %

\@pvspace{4.0pt}%
\@x{\makebox[10pt][r]{\scriptsize 6\hspace{0.8em}} chosenAt(b, v) \.{\defeq} \E\, Q \.{\in} \squorum \.{:} \A\, A \.{\in} Q \.{:} msg2b(
 A , b , v )}
 
\@pvspace{4.0pt}%
\@x{\makebox[10pt][r]{\scriptsize 7\hspace{0.8em}} chosen(v) \.{\defeq} \E\, B \.{\in} \sballot \.{:} chosenAt(B , v )}
 
\@pvspace{4.0pt}%
\begin{shaded3}
\@x{\makebox[10pt][r]{\scriptsize 8\hspace{0.8em}} showsSafeAtPaxos ( q , b , v ) \.{\defeq}}%
\@x{\makebox[10pt][r]{\scriptsize \hspace{0.8em}} \.{\land} \@s{1} \A\, A \.{\in} q \.{:} \E\, M_b \.{\in} \sballot \.{:} \E\, M_v \.{\in} \svalue \.{:} msg1b(A , b , M_b , M_v)}%
\@x{\makebox[10pt][r]{\scriptsize \hspace{0.8em}} \.{\land} \@s{2} \.{\lor} \@s{1} \A\, A \.{\in} \sacceptor \.{:} \A\, M_b \.{\in} \sballot \.{:} \A\, M_v \.{\in} \svalue \.{:} }
\@x{\@s{50} \neg (\, A \.{\in} q \@s{2} \.{\land} \@s{2} msg1b(A , b , M_b , M_v) \@s{2} \.{\land} \@s{2} (M_b \.{\neq} -1) \,)}%
\@x{\@s{21} \.{\lor} \@s{1} \E\, M_b \.{\in} \sballot \.{:} } \@x{\@s{33} \.{\land} \@s{2} \E\, A \.{\in} q \.{:} msg1b(A , b , M_b , v) \@s{2} \.{\land} \@s{2} (M_b \.{\neq} -1) }
\@x{\@s{33} \.{\land} \@s{2} \A\, A \.{\in} q \.{:} \A\, M_{b2} \.{\in} \sballot \.{:} \A\, M_{v2} \.{\in} \svalue \.{:} } \@x{\@s{48} msg1b(A , b , M_{b2} , M_{v2}) \@s{2} \.{\land} \@s{2} (M_{b2} \.{\neq} -1) \.{\rightarrow} M_{b2} \leq M_b}
\end{shaded3}

\@pvspace{4.0pt}%
\@x{\makebox[10pt][r]{\scriptsize 9\hspace{0.8em}} isSafeAtPaxos(b , v) \.{\defeq} \E\, Q {\in} \squorum \.{:} showsSafeAtPaxos(Q , b , v )}

\@pvspace{4.0pt}%
\@x{\makebox[10pt][r]{\scriptsize 10\hspace{0.8em}} Phase1a( b ) \@s{2} \.{\defeq}}
\@x{\makebox[10pt][r]{\scriptsize \hspace{0.8em}} \.{\land} \@s{2} b \.{\neq} -1}
\@x{\makebox[10pt][r]{\scriptsize \hspace{0.8em}} \.{\land} \@s{2} msg1a \.{'} \.{=} [ msg1a {\EXCEPT} {\bang} [ b ] \.{=} \top ]}%
\@x{\makebox[10pt][r]{\scriptsize \hspace{0.8em}} \.{\land} \@s{2} {\UNCHANGED}  msg1b , msg2a , msg2b , maxBal , maxVBal , maxVal}%

\@pvspace{4.0pt}%
\begin{shaded1}
\@x{\makebox[10pt][r]{\scriptsize 11\hspace{0.8em}} Phase1b( a , b ) \@s{2} \.{\defeq}}
\@x{\makebox[10pt][r]{\scriptsize \hspace{0.8em}} \.{\land} \@s{2} b \.{\neq} -1 \@s{3} \.{\land} \@s{3} msg1a(b) \@s{3} \.{\land} \@s{3} b > maxBal(a)}
\@x{\makebox[10pt][r]{\scriptsize \hspace{0.8em}} \.{\land} \@s{2} maxBal \.{'} \.{=} [ maxBal {\EXCEPT} {\bang} [ a ] \.{=} b ]}%
\@x{\makebox[10pt][r]{\scriptsize \hspace{0.8em}} \.{\land} \@s{2} msg1b \.{'} \.{=} [ msg1b {\EXCEPT} {\bang} [a, b, maxVBal(a), maxVal(a)] \.{=} \top ]}%
\@x{\makebox[10pt][r]{\scriptsize \hspace{0.8em}} \.{\land} \@s{2} {\UNCHANGED} msg1a , msg2a , msg2b , maxVBal , maxVal}%
\end{shaded1}

\@pvspace{4.0pt}%
\@x{\makebox[10pt][r]{\scriptsize 12\hspace{0.8em}} Phase2a( b, v ) \@s{2} \.{\defeq}}
\@x{\makebox[10pt][r]{\scriptsize \hspace{0.8em}} \.{\land} \@s{2} b \.{\neq} -1 \@s{3} \.{\land} \@s{3} v \.{\neq} none \@s{3} \.{\land} \@s{3} \neg(\, \E\, V \.{\in} \svalue \.{:} msg2a( b , V ) \,)}
\@x{\makebox[10pt][r]{\scriptsize \hspace{0.8em}} \.{\land} \@s{2} isSafeAtPaxos(b, v)}
\@x{\makebox[10pt][r]{\scriptsize \hspace{0.8em}} \.{\land} \@s{2} msg2a \.{'} \.{=} [ msg2a {\EXCEPT} {\bang} [b, v] \.{=} \top ]}%
\@x{\makebox[10pt][r]{\scriptsize \hspace{0.8em}} \.{\land} \@s{2} {\UNCHANGED} msg1a , msg1b , msg2b , maxBal , maxVBal , maxVal}%

\@pvspace{4.0pt}%
\@x{\makebox[10pt][r]{\scriptsize 13\hspace{0.8em}} Phase2b( a , b , v ) \@s{2} \.{\defeq}}
\@x{\makebox[10pt][r]{\scriptsize \hspace{0.8em}} \.{\land} \@s{2} b \.{\neq} -1 \@s{3} \.{\land} \@s{3} v \.{\neq} none \@s{3}\.{\land} \@s{3} msg2a(b, v) \@s{3} \.{\land} \@s{3} b \.{\geq} maxBal(a)}
\@x{\makebox[10pt][r]{\scriptsize \hspace{0.8em}} \.{\land} \@s{2} maxBal \.{'} \@s{6} \.{=} [ maxBal {\EXCEPT} {\bang} [ a ] \.{=} b ]}%
\@x{\makebox[10pt][r]{\scriptsize \hspace{0.8em}} \.{\land} \@s{2} maxVBal \.{'} \.{=} [ maxVBal {\EXCEPT} {\bang} [ a ] \.{=} b ]}%
\@x{\makebox[10pt][r]{\scriptsize \hspace{0.8em}} \.{\land} \@s{2} maxVal \.{'} \@s{7} \.{=} [ maxVal {\EXCEPT} {\bang} [ a ] \.{=} v ]}%
\@x{\makebox[10pt][r]{\scriptsize \hspace{0.8em}} \.{\land} \@s{2} msg2b \.{'} \@s{12} \.{=} [ msg2b {\EXCEPT} {\bang} [a, b, v] \.{=} \top ]}%
\@x{\makebox[10pt][r]{\scriptsize \hspace{0.8em}} \.{\land} \@s{2} {\UNCHANGED} msg1a , msg1b , msg2a}%

\@pvspace{4.0pt}%
\@x{\makebox[10pt][r]{\scriptsize 14\hspace{0.8em}} \init \.{\defeq} \.\A\, A \.{\in} \sacceptor \.{:} B \.{\in} \sballot \.{:} }
\@x{\@s{43} \.{\land} \@s{2} {\neg} msg1a(B)}
\@x{\@s{43} \.{\land} \@s{2} \.\A\, M_b \.{\in} \sballot \.{:} M_v \.{\in} \svalue \.{:} {\neg} msg1b(A, B, M_b, M_v)}
\@x{\@s{43} \.{\land} \@s{2} \.\A\, V \.{\in} \svalue \.{:} {\neg} msg2a(B, V) \@s{3} \.{\land} \@s{3} {\neg} msg2b(A, B, V)}
\@x{\@s{43} \.{\land} \@s{2} maxBal(A) = -1} %
\@x{\@s{43} \.{\land} \@s{2} maxVBal(A) = -1 \@s{2} \.{\land} \@s{2} maxVal(A) = none} %

\@pvspace{4.0pt}%
\@x{\makebox[10pt][r]{\scriptsize 15\hspace{0.8em}} \nex \.{\defeq} \.\E\, A \.{\in} \sacceptor \.{:} B \.{\in} \sballot \.{:} V \.{\in} \svalue \.{:} }
\@x{\@s{43} \.{\lor} \@s{2} Phase1a(B) \@s{15} \.{\lor} \@s{3} Phase1b(A, B)}
\@x{\@s{43} \.{\lor} \@s{2} Phase2a(B, V) \@s{3} \.{\lor} \@s{3} Phase2b(A, B, V)}

\@pvspace{4.0pt}%
 \@x{\makebox[10pt][r]{\scriptsize 16\hspace{0.8em}} \resizebox{0.95\linewidth}{!}{$\prop \@s{0} \.{\defeq} \@s{0} \.\A\, V_1, V_2 {\in} \svalue \.{:} chosen(V_1) \.{\land} chosen(V_2) \.{\rightarrow} V_1 \.{=} V_2$}}%

\@pvspace{4.0pt}%
\@x{\makebox[5pt][r]{\scriptsize \hspace{1em}}}\bottombar\@xx{\makebox[5pt][r]{\scriptsize \hspace{1em}}}%
\captionof{figure}{Lamport's \paxos protocol in pretty-printed TLA+}
\label{fig:paxos}
\end{algorithm}%

\subsection{Lamport's \voting Protocol}
\label{subsec:voting}
\noindent 
Figure~\ref{fig:voting} presents the TLA+~\cite{lamport2002specifying} description\footnote{Lamport's TLA+ encoding uses sets to denote variables. For example in~\cite{lamport_voting}, $votes[a]$ represents the set of votes cast by acceptor $a$. Throughout this paper, we use an equivalent representation based on relations/functions to enable encoding for SMT solving. $\langle b, v \rangle \in votes[a]$ is equivalently encoded in relational form as $votes(a, b, v) = \top$.} of the \voting protocol~\cite{lamport_voting}, which is a very high-level abstraction of \paxos that formalizes the way Lamport first thought about the Paxos consensus algorithm without getting distracted by details introduced by having the processes communicate by messages.
\voting has three unordered sorts named {$\svalue$, $\sacceptor$ and $\squorum$}, and a totally-ordered sort named $\sballot$. The protocol has two state symbols, $votes$ and $maxBal$ defined on these sorts that serve as the protocol's state variables. $votes(a,b,v)$ is true iff an acceptor $a$ has voted for value $v$ in ballot number $b$. $maxBal(a)$ returns a ballot number such that acceptor $a$ will never cast any further vote in a ballot numbered less than $maxBal(a)$. The global axiom (line 5) defines the elements of the $\squorum$ sort to be subsets of the $\sacceptor$ sort and restricts them further by requiring them to be pair-wise non-disjoint. Lines 6-9 specify definitions $chosenAt$, $chosen$, $showsSafeAt$, and $isSafeAt$, which serve as auxiliary non-state variables. 
Protocol transitions are specified by the actions $IncreaseMaxBal$ and $VoteFor$ (lines 10-11), and lines 12-14 specify the protocol's initial states, transition relation, and safety property.

Viewed as a parameterized system, the template of the \voting protocol is $\voting$($\svalue$, $\sacceptor$, $\squorum$, $\sballot$).
Its finite instance:
{
\eqsize
\begin{align}
& \voting(2,3,3,4) : \nonumber\\
& \hspace{15pt} \fin{value_2} \triangleq \{ \fin{v_1}, \fin{v_2} \} \nonumber\\
& \hspace{15pt} \fin{acceptor_3} \triangleq \{ \fin{a_1, a_2, a_3} \} \nonumber\\
& \hspace{15pt} \fin{quorum_3} \triangleq \{ \fin{q_{12}\!:\!\{a_1,a_2\},~q_{13}\!:\!\{a_1,a_3\},~q_{23}\!:\!\{a_2,a_3\}} \} \nonumber\\
& \hspace{15pt} \fin{ballot_4} \triangleq [ \fin{b_{min}, b_1, b_2, b_{max}} ] \nonumber
\end{align}
}%
has three finite symmetric sorts named $\fin{value_2}$, $\fin{acceptor_3}$ and $\fin{quorum_3}$, defined as sets of arbitrarily-named distinct constants, while the finite totally-ordered sort $\fin{ballot_4}$ is composed of a list of ordered constants, i.e., $\fin{b_{min} < b_1 < b_2 < b_{max}}$, where $\fin{b_{min}} = -1$ since $-1$ is the ``minimum'' ballot number. The constants of the $\fin{quorum_3}$ sort are subsets of the $\fin{acceptor_3}$ sort and are named to reflect their symmetric dependence on the $\fin{acceptor_3}$ sort. This instance has $24$ $votes$ state variables that return a $\BOOLEAN$ and $3$ $maxBal$ state variables that return a ballot number in $\fin{ballot_4}$. A \textit{state} of this instance corresponds to a complete assignment to these $27$ state variables.

\subsection{Lamport's \paxos Protocol}
\label{subsec:paxos}
\noindent
Figure~\ref{fig:paxos} presents the TLA+ description of Lamport's \paxos protocol~\cite{lamport_paxos}, which is a specification of the Paxos consensus algorithm~\cite{lamport1998part,lamport2001paxos}. \paxos implements \voting through the refinement mapping $[votes \leftarrow msg2b$, $maxBal \leftarrow maxBal]$, where acceptors now communicate with each other through distributed message passing. State variables $msg1a$, $msg1b$, $msg2a$, and $msg2b$ are used to model the set of different messages that can be sent in the protocol, corresponding to actions $Phase1a$, $Phase1b$, $Phase2a$, and $Phase2b$ respectively. The pair $\langle maxVBal(a), maxVal(a) \rangle$ is the vote with the largest ballot number cast by acceptor $a$. The ballot $b$ leader can send a $msg1a(b)$ by performing the action $Phase1a(b)$. $Phase1b(a, b)$ implements the $IncreaseMaxBal(a, b)$ action from \voting, where after receiving $msg1a(b)$, acceptor $a$ sends $msg1b$ to the ballot $b$ leader containing the values of $maxVBal(a)$ and $maxVal(a)$. In the $Phase2a(b, v)$ action, the ballot $b$ leader sends $msg2a$ asking the acceptors to vote for a value $v$ that is safe at ballot number $b$. Its enabling condition $isSafeAtPaxos(b, v)$ checks the enabling condition $isSafeAt(b, v)$ from \voting. $Phase2b$ implements the $VoteFor$ action in \voting, and enables acceptor $a$ to vote for value $v$ in ballot number $b$. We refer the reader to~\cite{lamport_turing_video} for a detailed explanation to understand the internals of \paxos.

Represented as a parameterized system $\paxos$($\svalue$, $\sacceptor$, $\squorum$, $\sballot$), its finite instance $\paxos(2,3,3,4)$ has $132$ $\BOOLEAN$ state variables, $6$ state variables that return a ballot number in $\fin{ballot_4}$, and $3$ state variables that return a value in $\fin{value_2}$.

\subsection{Intermediate Levels between \voting and \paxos}
\label{subsec:paxos_intermediate}
\noindent
We introduced two intermediate levels, \paxossimple and \paxosabstract, between \voting and \paxos (Appendix~\ref{app:paxos_intermediate}). These intermediate levels are abstractions of \paxos, inspired from the already-existing literature~\cite{lamport2005generalized,peluso2016making,padon2017paxos,multipaxos_tla,ivy_paxos2}.
\paxosabstract is inspired from the specification of Generalized Paxos by Lamport~\cite{lamport2005generalized} and uses a commonly-used encoding transformation, as utilized in~\cite{multipaxos_tla,padon2017paxos,ivy_paxos2}. Instead of explicitly keeping a track of $maxVBal(a)$ and $maxVal(a)$, \paxosabstract abstracts them away and implicitly computes their respective values using the history of all votes cast by the acceptor $a$, i.e., using the history of $msg2b$ from acceptor $a$, by modifying the $Phase1b(a, b)$ action (line 11 in Figure~\ref{fig:paxos}) to as shown in Figure~\ref{fig:paxos_implicit}.

\begin{algorithm}[!tb]
\begin{shaded2}
{
\batchmode
\tlatex
\tlasize
\setboolean{shading}{true}

\@x{\makebox[5pt][r]{\scriptsize \hspace{1em}}}
\@s{60} \@xx{ {\MODULE} \text{\paxosabstract}} \@xx{\makebox[5pt][r]{\scriptsize \hspace{1em}}}%
\@pvspace{4.0pt}%

\@x{ \makebox[5pt][r]{\scriptsize 11\hspace{0.8em}} Phase1b( a , b ) \@s{2} \.{\defeq}}
\@x{ \@s{5} \.{\land} \@s{2} b \.{\neq} -1 \@s{3} \.{\land} \@s{3} msg1a(b) \@s{3} \.{\land} \@s{3} b > maxBal(a)}
\@x{ \@s{5} \.{\land} \@s{2} maxBal \.{'} \.{=} [ maxBal {\EXCEPT} {\bang} [ a ] \.{=} b ]}%
\@x{ \@s{5} \.{\land} \@s{2} \E\, M_b \.{\in} \sballot \.{:}  \E\, M_v \.{\in} \svalue \.{:} }
\@x{ \@s{5} \@s{6} \.{\land} \@s{2} \.{\lor} \@s{2} \.{\land} \@s{2} (M_b = -1)}
\@x{ \@s{5} \@s{32} \.{\land} \@s{2} \A\, B \.{\in} \sballot \.{:} \A\, V \.{\in} \svalue \.{:} \neg msg2b(a , B, V)}
\@x{ \@s{5} \@s{19} \.{\lor} \@s{2} \.{\land} \@s{2} (M_b \neq -1) \@s{2} \.{\land} \@s{2} msg2b(a, M_b, M_v)}
\@x{ \@s{5} \@s{32} \.{\land} \@s{2} \A\, B \.{\in} \sballot \.{:} \A\, V \.{\in} \svalue \.{:}}
\@x{ \@s{5} \@s{108} msb2b(a, B, V) \.{\rightarrow} B \leq M_b}

\@x{ \@s{5} \@s{6} \.{\land} \@s{2} msg1b \.{'} \.{=} [ msg1b {\EXCEPT} {\bang} [a, b, M_b, M_v] \.{=} \top ]}%
\@x{ \@s{5} \.{\land} \@s{2} {\UNCHANGED} msg1a , msg2a , msg2b}%

}%
\end{shaded2}
\captionof{figure}{Modifications in \paxosabstract compared to \paxos}
\label{fig:paxos_implicit}
\end{algorithm}%

\paxossimple further simplifies \paxosabstract and eliminates tracking of the maximum ballot (and the corresponding value) in which an acceptor voted from $msg1b$ completely, i.e., the last two arguments of $msg1b$ are abstracted away. Instead, the history of all votes cast is used to describe how new votes are cast. This is done by replacing the definition $showsSafeAtPaxos$ (line 8 in Figure~\ref{fig:paxos}) with its simplified form, expressed using $msg2b$ as shown in Figure~\ref{fig:paxos_simple}.

\begin{algorithm}[!tb]
\begin{shaded4}
{
\batchmode
\tlatex
\tlasize
\setboolean{shading}{true}
\@x{\makebox[5pt][r]{\scriptsize \hspace{1em}}}
\@s{60} \@xx{ {\MODULE} \text{\paxossimple}} \@xx{\makebox[5pt][r]{\scriptsize \hspace{1em}}}%
\@pvspace{4.0pt}%

\@x{ \makebox[5pt][r]{\scriptsize 8\hspace{0.8em}} showsSafeAtSimplePaxos ( q , b , v ) \.{\defeq}}%
\@x{ \@s{5} \.{\land} \@s{1} \A\, A \.{\in} q \.{:} msg1b(A , b)}%
\@x{ \@s{5} \.{\land} \@s{2} \.{\lor} \@s{1} \A\, A \.{\in} \sacceptor \.{:} \A\, M_b \.{\in} \sballot \.{:} \A\, M_v \.{\in} \svalue \.{:} }
\@x{ \@s{5} \@s{40} \neg (\, A \.{\in} q \@s{2} \.{\land} \@s{2} msg1b(A , b) \@s{2} \.{\land} \@s{2} msg2b(A, M_b, M_v) \,)}%
\@x{ \@s{5} \@s{13} \.{\lor} \@s{1} \E\, M_b \.{\in} \sballot \.{:} } 
\@x{ \@s{5} \@s{23} \.{\land} \@s{2} \E\, A \.{\in} q \.{:} msg1b(A , b) \@s{2} \.{\land} \@s{2} msg2b(A, M_b, v) }
\@x{ \@s{5} \@s{23} \.{\land} \@s{2} \A\, A \.{\in} q \.{:} \A\, M_{b2} \.{\in} \sballot \.{:} \A\, M_{v2} \.{\in} \svalue \.{:} } 
\@x{ \@s{5} \@s{38} msg1b(A , b) \@s{2} \.{\land} \@s{2} msg2b(A, M_{b2}, M_{v2}) \.{\rightarrow} M_{b2} \leq M_b}
}%
\end{shaded4}
\captionof{figure}{Modifications in \paxossimple compared to \paxosabstract}
\label{fig:paxos_simple}
\end{algorithm}%

\section{Hierarchical Verification of Paxos}
\label{sec:paxos_hr}
\noindent
Using the 4-level hierarchy $\paxos \prec \paxosabstract \prec \paxossimple \prec \voting$, this section is a ``log'' of how IC3PO automatically derived the required strengthening assertions that established the safety of \textit{Paxos}.

\subsection{Proving \voting}
\noindent
Using instance $\voting(2,3,3,4)$, \icpo proved the safety of \voting by automatically deriving the inductive invariant $\ev inv \triangleq \ev Safety \wedge \ev A_1 \wedge \ev A_2$ where
{
\asize
\begin{align}
& \ev A_1 =~ \forall A \in \sacceptor, B \in \sballot, V \in \svalue : \nonumber\\
& \hspace{40pt} votes(A, B, V) \to isSafeAt(B, V) \nonumber\\
& \ev A_2 =~ \forall A \in \sacceptor, B \in \sballot, V_1, V_2 \in \svalue : \nonumber\\
& \hspace{37pt} chosenAt(B, V_1) \wedge votes(A, B, V_2) \to (V_1 = V_2) \nonumber
\end{align}
}%
In words, these two strengthening assertions mean:
\begin{shaded}
\begin{itemize}
    \item[$A_1$:] If an acceptor voted for value $V$ in ballot number $B$, then $V$ is safe at $B$.
    \item[$A_2$:] If value $V_1$ is chosen at ballot $B$, then no acceptor can vote for a value different than $V_1$ in $B$.
\end{itemize}
\end{shaded}

\subsection{Proving \paxossimple}
\noindent
Using the refinement mapping $[votes \leftarrow msg2b$, $maxBal \leftarrow maxBal]$, \icpo transformed $\ev A_1$ and $\ev A_2$ to the following corresponding versions for \paxossimple:
{
\asize
\begin{align}
& \eps A_1 =~ \forall A \in \sacceptor, B \in \sballot, V \in \svalue : \nonumber\\
& \hspace{40pt} msg2b(A, B, V) \to isSafeAt(B, V) \nonumber\\
& \eps A_2 =~ \forall A \in \sacceptor, B \in \sballot, V_1, V_2 \in \svalue : \nonumber\\
& \hspace{40pt} chosenAt(B, V_1) \wedge msg2b(A, B, V_2) \to (V_1 = V_2) \nonumber
\end{align}
}%
These two assertions, passed down from the proof of \voting, represented a strengthening of the safety property of \paxossimple that allowed \icpo to prove it with the inductive  invariant $\eps inv \triangleq \eps Safety \wedge \mathop  \bigwedge \nolimits_{1 \le i \le 6} \eps A_i$ where  
{
\asize
\begin{align}
& \eps A_3 =~ \forall B \in \sballot, V \in \svalue : \nonumber\\
& \hspace{50pt} msg2a(B, V) \to isSafeAt(B, V) \nonumber\\
& \eps A_4 =~ \forall B \in \sballot, V_1, V_2 \in \svalue : \nonumber\\
& \hspace{50pt} msg2a(B, V_1) \wedge msg2a(B, V_2) \to (V_1 = V_2) \nonumber\\
& \eps A_5 =~ \forall A \in \sacceptor, B \in \sballot, V \in \svalue : \nonumber\\
& \hspace{50pt} msg2b(A, B, V) \to msg2a(B, V) \nonumber\\
& \eps A_6 =~ \forall A \in \sacceptor, B \in \sballot : \nonumber\\
& \hspace{50pt} msg1b(A, B) \to maxBal(A) \geq B \nonumber
\end{align}
}%
are four additional automatically-generated strengthening assertions that express the following facts about \paxossimple:
\begin{shaded}
\begin{itemize}
    \item[$A_3$:] If ballot $B$ leader sends a $2a$ message for value $V$, then $V$ is safe at $B$.
    \item[$A_4$:] A ballot leader can send $2a$ messages only for a unique value.
    \item[$A_5$:] If an acceptor voted for a value in ballot number $B$, then there is a $2a$ message for that value at $B$.
    \item[$A_6$:] If an acceptor has sent a $1b$ message at a ballot number $B$, then its $maxBal$ is at least as high as $B$.
\end{itemize}
\end{shaded}

\subsection{Proving \paxosabstract}
\noindent
All variables from \paxossimple refine to \paxosabstract as is, except for $msg1b$ that adds explicit tracking of the maximum vote voted by an acceptor in \paxosabstract. 
Assertions $\eps A_1$ to $\eps A_5$ map to $\epa A_1$ to $\epa A_5$ in \paxosabstract as is, while $\eps A_6$ maps as:
{
\asize
\begin{align}
& \epa A_6 =~ \forall A \in \sacceptor, B, B_{max} \in \sballot, V_{max} \in \svalue : \nonumber\\
& \hspace{50pt} msg1b(A, B, B_{max}, V_{max}) \to maxBal(A) \geq B \nonumber
\end{align}
}%
These six assertions, passed down from the proof of \paxossimple, represented a strengthening of the safety property of \paxosabstract that allowed \icpo to prove it with the inductive  invariant $\epa inv \triangleq \epa Safety \wedge \mathop  \bigwedge \nolimits_{1 \le i \le 8} \epa A_i$ where  
{
\asize
\begin{align}
& \epa A_7 =~ \forall A \in \sacceptor, B, B_{max} \in \sballot, V_{max} \in \svalue : \nonumber\\
& \hspace{16pt} [(B > -1) \wedge (B_{max} > -1) \wedge msg1b(A, B, B_{max}, V_{max})] \nonumber\\
& \hspace{115pt} \to msg2b(A, B_{max}, V_{max}) \nonumber\\
& \epa A_8 = \forall A \in \sacceptor, B, B_{mid}, B_{max} \in \sballot, \nonumber\\
& \hspace{37pt} V, V_{max} \in \svalue : \nonumber\\
& \hspace{10pt} [(B > B_{mid}) \wedge (B_{mid} > B_{max}) \wedge msg1b(A, B, B_{max}, V_{max})] \nonumber\\
& \hspace{115pt} \to \neg msg2b(A, B_{mid}, V) \nonumber
\end{align}
}%
are two additional automatically-generated strengthening assertions that express the following facts about \paxosabstract:
\begin{shaded}
\begin{itemize}
    \item[$A_7$:] If an acceptor issued a $1b$ message at ballot number $B$ with the maximum vote $\langle B_{max}, V_{max} \rangle$, and both $B$ and $B_{max}$ are higher than $-1$, then  the acceptor has voted for value $V_{max}$ in ballot $B_{max}$.
    \item[$A_8$:] If an acceptor issued a $1b$ message at ballot number $B$ with the maximum vote $\langle B_{max}, V_{max} \rangle$, then  the acceptor cannot have voted in any ballot number strictly between $B_{max}$ and $B$.
\end{itemize}
\end{shaded}

\subsection{Proving \paxos}
\noindent
All variables from \paxosabstract refine to \paxos trivially, mapping $\epa A_1,\dots,\epa A_8$ to $\ep A_1,\dots,\ep A_6$ in \paxos as is.
These eight assertions, passed down from the proof of \paxosabstract, represented a strengthening of the safety property of \paxos that allowed \icpo to prove it with the inductive  invariant $\ep inv \triangleq \ep Safety \wedge \mathop  \bigwedge \nolimits_{1 \le i \le 11} \ep A_i$ where  
{
\asize
\begin{align}
& \ep A_9 =~ \forall A \in \sacceptor : maxVBal(A) \leq maxBal(A) \nonumber\\
& \ep A_{10} =~ \forall A \in \sacceptor, B \in \sballot, V \in \svalue : \nonumber\\
& \hspace{60pt} msg2b(A, B, V) \to maxVBal(A) \geq B \nonumber\\
& \ep A_{11} =~ \forall A \in \sacceptor : maxVBal(A) > -1 \nonumber\\
& \hspace{60pt} \to msg2b(A, maxVBal(A), maxVal(A)) \nonumber
\end{align}
}%
are three additional automatically-generated strengthening assertions that express the following facts about \paxos:
\begin{shaded}
\begin{itemize}
    \item[$A_9$:] $maxVBal$ of an acceptor is less than or equal to its $maxBal$.
    \item[$A_{10}$:] If an acceptor voted in a ballot number $B$, then its $maxVBal$ is at least as high as $B$.
    \item[$A_{11}$:] If acceptor $A$ has its $maxVBal$ higher than $-1$, then $A$ has already cast a vote $\langle maxVBal(A), maxVal(A)\rangle$.
\end{itemize}
\end{shaded}

\section{Discussion}
\label{sec:paxos_discuss}
\noindent
This section provides a discussion about certain key points and features about the \paxos proof from Section~\ref{sec:paxos_hr}.

\subsection{Comparison against Human-written Invariants}
\noindent
Optionally, the inductive invariant $\ep inv$ can be minimized to derive a subsumption-free and closed set of invariants, which removes $A_1$ and $A_2$ that are subsumed by the conjunction $A_3 \wedge A_4 \wedge A_5$.
After this minimization, the inductive invariant of \paxos matches identically with the manually-written and TLAPS-checked inductive invariant from~\cite{paxos_proof}, guaranteeing its correctness.
Similarly, the inductive invariant of \voting, i.e., $\ev inv$, matches directly with the manually-written and TLAPS-checked inductive invariant from~\cite{voting_proof}.

\subsection{Benefits of Range Boosting}
\noindent
Assertions $A_6$ to $A_{11}$ express conditions defined over ordered ranges in the \textit{infinite} totally-ordered $\sballot$ domain. Inferring such invariants automatically through \icpo becomes possible through range boosting (Section~\ref{sec:range}), that extends incremental induction with the knowledge of \textit{temporal regularity} over totally-ordered domains by learning quantified clauses over ordered ranges.

\subsection{Protocol's Formula Structure}
\noindent
Note that $A_1$ to $A_3$ use definitions $isSafeAt$ and $chosenAt$, which implicitly enables \icpo to incorporate learning with complex quantifier alternations. Inspired from previous works on the importance of using derived/ghost variables~\cite{lamport1977proving,owicki1976verifying,namjoshi2007symmetry}, \icpo utilizes the \textit{formula structure} of the protocol's transition relation in a unique manner, by incorporating \textit{definitions} in the protocol specification as auxiliary non-state variables during reachability analysis, described in detail in~\cite{goel2021on}. This provides a simple and inexpensive procedure to incorporate clause learning with complex quantifier alternations.

\subsection{Decidability}
\noindent
Protocol specifications at each of the four levels include quantifier alternation cycles that make unbounded SMT reasoning fall into the undecidable fragment of first-order logic.
Unsurprisingly, previous works that rely on unbounded SMT reasoning, like SWISS~\cite{hance2021finding}, fol-ic3~\cite{koenig2020first}, DistAI~\cite{yao2021distai}, I4~\cite{ma2019i4}, and UPDR~\cite{karbyshev2017property}, struggle with verifying Lamport's Paxos. \icpo, on the other hand, performs incremental induction and finite convergence over finite protocol instances using finite-domain reasoning that is always decidable.

\subsection{Why a Four-Level Hierarchy?}
\noindent
The original Paxos specification is composed of a two-level hierarchy $\paxos \prec \voting$. 
Given the two strengthening assertions $A_1$ and $A_2$ from \voting, inferring the remaining nine assertions for \paxos directly in one step of hierarchical strengthening is difficult, since these two specifications are too far apart to be proved directly. \icpo struggled with the large state space of \paxos and learnt too many weak clauses involving $msg1b$, $maxVBal$ and $maxVal$, eventually running out of memory due to invariant inference getting confused with several counterexamples-to-induction. 
Table~\ref{tab:state_space} compares the state-space size of protocol instances at each of the four hierarchical levels.
Even though $2^{147}$ is not huge, especially with respect to hardware verification problems~\cite{hwmcc2020,goel2019empirical,goel2019model}, \paxos has a dense state-transition graph where state-transitions are tightly coupled with high in- and out- degree, making the problem difficult for automatic invariant inference with incremental induction based model checking.

\begin{table}[!b]
\centering
\renewcommand{\arraystretch}{1.1}
\begin{tabular}{l|c}
\hline
\multicolumn{1}{c|}{\textbf{Finite Instance}} & \textbf{State-space Size}  \\
\hline
$\voting(2,3,3,4)$ & $2^{30}$ \\
$\paxossimple(2,3,3,4)$ & $2^{54}$ \\
$\paxosabstract(2,3,3,4)$ & $2^{138}$ \\
$\paxos(2,3,3,4)$ & $2^{147}$ \\
\hline
\end{tabular}
\caption{State-space size for finite instances with 2 $\fin{value}$, 3 $\fin{acceptor}$, 3 $\fin{quorum}$, and 4 $\fin{ballot}$}
\label{tab:state_space}
\end{table}

Adding \paxosabstract reduced the complexity in \paxos by abstracting away $maxVBal$ and $maxVal$. 
Still, scalability remained a challenge due to $msg1b$, that contributed to 96 out of 147 state bits in $\paxos(2,3,3,4)$.
Adding another level, i.e., $\paxossimple$, removed 84 out of these 96 state bits by abstracting away explicit tracking of the maximum vote of an acceptor from $msg1b$.
When compared against $\paxos$, $\paxossimple$ is significantly simpler, with a total state-space size to be just $2^{54}$ for its finite instance $\paxossimple(2,3,3,4)$, which led \icpo to successfully prove \paxos automatically using the four-level hierarchy.

\subsection{Extension to \multipaxos and \flexiblepaxos}
\noindent 
Till now, by \paxos we meant \textit{single-decree} Paxos which is the core consensus algorithm underlying the complete Paxos state-machine replication protocol~\cite{lamport1998part,lamport2001paxos}, commonly referred to as \multipaxos~\cite{multipaxos_tla}. In \multipaxos, a sequence of instances execute single-decree \paxos such that the value chosen in the $i^{th}$ instance becomes the $i^{th}$ command executed by the replicated state machine. Additionally, if the leader is relatively stable, $Phase1$ becomes unnecessary and is skipped, reducing the failure-free message delay from 4 delays to 2 delays.

Mapping each of the assertions $A_1,\dots,A_{11}$ to \multipaxos is trivial, and simply adds the corresponding instance as an additional universally-quantified argument, e.g., $A_{11}$ maps as:
{
\asize
\begin{align}
& \emp A_{11} = \forall A \in \sacceptor, I \in \sinstance : \nonumber\\
& \hspace{50pt} maxVBal(A, I) > -1 \nonumber\\
& \hspace{50pt} \to msg2b(A, I, maxVBal(A, I), maxVal(A, I)) \nonumber
\end{align}
}%
Unsurprisingly, the 11 strengthening assertions, passed down from the proof of \paxos, together with the safety property of \multipaxos, allowed \icpo to trivially prove it with no additional strengthening assertions needed, meaning $\emp \prop \wedge \mathop  \bigwedge \nolimits_{1 \le i \le 11} \emp A_i$ is already an inductive invariant of \multipaxos. As described in previous works~\cite{de2000revisiting,lamport1998part,lamport2001paxos,chand2016formal}, the crux of proving the safety of \multipaxos is based on proving single-decree \paxos since each consensus instance participates independently without any interference from other instances. Our experiments validated this further.

Similarly, we also tried another Paxos variant called \flexiblepaxos~\cite{DBLP:journals/corr/HowardMS16}, which also verifies trivially with the same inductive invariant, i.e., with no additional strengthening assertions needed.

\section{Experiments}
\label{sec:exp}
\begin{table*}[!tb]
\small
\setlength\tabcolsep{4.5pt}
\begin{center}
\resizebox{\textwidth}{!}{
\begin{tabular}{ll|c|rccccc|cc|rr}
    \multicolumn{2}{c}{} &  \multicolumn{1}{c|}{} & \multicolumn{6}{c|}{Time (seconds)} & \multicolumn{2}{c|}{Inv} & \multicolumn{2}{c}{SMT} \\
    \hline
    \multicolumn{2}{c|}{Protocol} & \multicolumn{1}{c|}{S.A.} & \multicolumn{1}{c}{\icpo} & 
    \multicolumn{1}{c}{\swiss} & \multicolumn{1}{c}{\folic} & \multicolumn{1}{c}{\distai} & \multicolumn{1}{c}{\ifour} & \multicolumn{1}{c|}{\updr} & \multicolumn{1}{c}{\icpo} & \multicolumn{1}{c|}{Human} & \multicolumn{1}{c}{\icpo} & \multicolumn{1}{c}{\ifour} \\
	\hline
	\multirow{3}{*}{\rotatebox[origin=c]{90}{\epr}} & epr-paxos  & $\varnothing$ & \textbf{568} & 15950\textcolor{failcolor}{$^*$} & \textcolor{failcolor}{timeout} & \textcolor{failcolor}{error} & \textcolor{failcolor}{memout} & \textcolor{failcolor}{timeout} & 6 & 11 & 5680 & 1701556 \\
	& epr-flexible\_paxos  & $\varnothing$ & \textbf{561} & 18232\textcolor{failcolor}{$^*$} & \textcolor{failcolor}{timeout} & \textcolor{failcolor}{error} & \textcolor{failcolor}{memout} & \textcolor{failcolor}{failure} & 6 & 11 & 1509 & 1761504 \\
	& epr-multi\_paxos  & $\varnothing$ & \textcolor{failcolor}{timeout} & \textcolor{failcolor}{timeout} & \textcolor{failcolor}{timeout} & \textcolor{failcolor}{error} & \textcolor{failcolor}{memout} & \textcolor{failcolor}{timeout} & \textcolor{failcolor}{$-$} & 12 & \textcolor{failcolor}{$-$} & 1902621 \\
    \hline
	\multirow{6}{*}{\rotatebox[origin=c]{90}{\orig}} & \voting  & $\varnothing$ & \textbf{64} & \textcolor{failcolor}{timeout} & \textcolor{failcolor}{timeout} & \textcolor{failcolor}{error} & \textcolor{failcolor}{memout} & \textcolor{failcolor}{timeout} & 3 & 3 & 1057 & 1714170 \\
	& \paxossimple  & $A_{1-2}$ & \textbf{51} & \textcolor{failcolor}{timeout} & \textcolor{failcolor}{timeout} & \textcolor{failcolor}{error} & \textcolor{failcolor}{failure} & \textcolor{failcolor}{timeout} & 5 & 5 & 618 & 158470 \\
	& \paxosabstract  & $A_{1-6}$ & \textbf{2008} & \textcolor{failcolor}{timeout} & \textcolor{failcolor}{timeout} & \textcolor{failcolor}{error} & \textcolor{failcolor}{failure} & \textcolor{failcolor}{timeout} & 7 & 7 & 18329 & 69715 \\
	& \paxos  & $A_{1-8}$ & \textbf{98} & \textcolor{failcolor}{timeout} & \textcolor{failcolor}{timeout} & \textcolor{failcolor}{error} & \textcolor{failcolor}{failure} & \textcolor{failcolor}{timeout} & 10 & 10 & 668 & 76030 \\
	& \multipaxos  & $A_{1-11}$ & \textbf{340} & \textcolor{failcolor}{timeout} & \textcolor{failcolor}{timeout} & \textcolor{failcolor}{error} & \textcolor{failcolor}{timeout} & \textcolor{failcolor}{timeout} & 10 & 10 & 161 & \textcolor{failcolor}{$-$} \\
	& \flexiblepaxos  & $A_{1-11}$ & \textbf{1408} & \textcolor{failcolor}{timeout} & \textcolor{failcolor}{timeout} & \textcolor{failcolor}{error} & \textcolor{failcolor}{failure} & \textcolor{failcolor}{timeout} & 10 & 10 & 161 & 6983 \\
	\hline
\end{tabular}
}
\captionsetup{justification=centering}
\captionsetup{belowskip=0pt}
\caption{Comparison of \icpo~against other state-of-the-art verifiers}
\captionsetup{justification=raggedright, aboveskip=0pt,belowskip=0pt}
\caption*{
\orig problems employ hierarchical strengthening (as detailed in Section~\ref{sec:paxos_hr}), while \epr problems do not.\\
Column 2 (labeled S.A.) lists strengthening assertions added through hierarchical strengthening to the safety property ($\varnothing$ means none).\\
Columns 3-8 (labeled Time) compare the runtime in seconds. For failed \swiss runs, we include the runtime from~\cite{hance2021finding} (indicated with \textcolor{failcolor}{$^*$}). \\
Columns 9-10 (labeled Inv) compare number of assertions in the inductive invariant between \icpo (with subsumption checking and minimization) and human-written proofs. \\
Columns 11-12 (labeled SMT) compare total number of SMT queries made by \icpo versus \ifour (until failure for unsuccessful runs).
}
\label{tab:table_exp}
\end{center}
\end{table*}

\noindent
\icpo~\cite{ic3pogithub} currently accepts protocol descriptions in the Ivy language~\cite{padon2016ivy} and uses the Ivy compiler to extract a logical formulation of the protocol in a SMT-LIB~\cite{BarFT-SMTLIB} compatible format.
To get an idea on the effectiveness of hierarchical strengthening, we also evaluated automatically deriving inductive proofs for EPR variants of Paxos from~\cite{padon2017paxos} without any hierarchical strengthening. These specifications describe Paxos in the EPR fragment~\cite{piskac2010deciding} of first-order logic and also incorporate simplifications equivalent to the ones described for \paxossimple in Section~\ref{subsec:paxos_intermediate}.
We performed a detailed comparison against other state-of-the-art techniques for automatically verifying distributed protocols:
\begin{itemize}
    \item[--] \swiss~\cite{hance2021finding} uses SMT solving to derive an inductive invariant by performing an enumerative search in an optimized and bounded invariant search space.
    \item[--] \folic~\cite{koenig2020first}, implemented in \textit{mypyvy}~\cite{mypyvygithub}, extends IC3 with a separators-based technique that performs enumerative search for a quantified separator in the space of bounded mixed quantifier prefixes.
    \item[--] \distai~\cite{yao2021distai} performs data-driven invariant learning by enumerating over possible invariants derived from simulating a protocol at different instance sizes, followed by iteratively refining and checking candidate invariants.
    \item[--] \ifour~\cite{ma2019i4,ma2019towards} performs finite-domain IC3 (without accounting for regularity) using the AVR model checker~\cite{goel2019model,avr2}, followed by iteratively generalizing and checking the inductive invariant produced by AVR.
    \item[--] \updr, from the \textit{mypyvy}~\cite{mypyvygithub} framework, implements PDR$^{\forall}$/UPDR~\cite{10.1145/3022187} for verifying distributed protocols.
\end{itemize}
All experiments were performed on an Intel (R) Xeon CPU (X5670). For each run, we used a 5-hour timeout and a 32 GB memory limit. All tools were executed in their respective default configurations. We used Z3~\cite{demoura2008z3} version 4.8.10, Yices 2~\cite{dutertre2014yices} version 2.6.2, and CVC4~\cite{BCD_11} version 1.8.

\subsection{Results}
\noindent Table~\ref{tab:table_exp} summarizes the experimental results. \epr variants were run without any hierarchical strengthening. For \orig problems, we employed hierarchical strengthening using each tool to verify Lamport's original Paxos specification (and its variants) through higher-level strengthening assertions that were automatically generated from \icpo (as detailed in Section~\ref{sec:paxos_hr}). Note that \orig problems include quantifier-alternation cycles that make unbounded SMT reasoning fall into the undecidable fragment of first-order logic. 

\icpo emerges as the only successful technique that verifies Lamport's Paxos and its variants, and automatically infers the required inductive invariants efficiently. Unsurprisingly, none of the other tools (i.e., \swiss, \folic, \distai, \ifour and \updr) were able to solve \orig problems since each of these tools rely on unbounded SMT reasoning and struggle on problems that fall outside the decidable EPR fragment of first-order logic.

\subsection{Discussion}

\subsubsection*{Effect of hierarchical strengthening}
Comparing \epr versus \orig shows the advantages offered by hierarchical strengthening.
Even though \icpo was able to automatically verify EPR versions of single-decree Paxos and flexible Paxos from~\cite{padon2017paxos}, none of the tools were able to automatically verify the EPR version of multi-decree Paxos. \orig variants, on the other hand, employed hierarchical strengthening which allowed \icpo to verify Lamport's Paxos automatically and efficiently by using the protocol's hierarchical structure.

\subsubsection*{Comparison against other verifiers}
\distai failed on all problems due to unsupported constructs and parsing errors.
\ifour and \updr (as well as \distai) are limited to generating only universally-quantified invariants over state variables, and hence, were unable to solve any problem.
While both \icpo and \ifour use incremental induction over a finite protocol instance, the number of SMT queries made by \ifour grows drastically, indicating the benefits offered by symmetry and range boosting employed in \icpo. 
\folic also fails on all problems, showing limited scalability of its enumeration-based separators technique operating directly in the unbounded domain. For \swiss, we weren't able to replicate results for \epr problems as reported in~\cite{hance2021finding} using our experimental setup. Nevertheless, \swiss showed limited capabilities for solving \orig problems.

\subsubsection*{Comparison against human-written invariants}
As evident from $A_1$ to $A_{11}$ in Section~\ref{sec:paxos_hr}, \icpo generated concise, human-readable inductive invariants. In fact, every invariant of \paxos written manually by Lamport et al. (as detailed in~\cite{lamport_paxos,paxos_proof}) had a corresponding equivalent invariant in the inductive proof automatically generated with \icpo. In contrast, deriving such invariants manually, even in the presence of a hierarchical structure, is a tedious and error-prone process that demands deep domain expertise~\cite{paxos_proof,hawblitzel2015ironfleet,padon2017paxos,taube2018modularity}. 

Overall, the evaluation confirms our main hypothesis, that it is possible to utilize the regularity and hierarchical structure in complex distributed protocols, like in Paxos, to scale automatic verification beyond the current state-of-the-art.

\section{Related Work}
\label{sec:related}
\noindent Introduced by Lamport, TLA+ is a widely-adopted language for the specification and verification of distributed protocols~\cite{newcombe2015amazon,beers2008pre}. The TLA+ toolbox~\cite{tla_toolbox} provides the TLC model checker, which is primarily used as a debugging tool for verifying small finite protocol instances~\cite{yu1999model}, and not as a tool for inferring inductive invariants. The TLAPS proof assistant~\cite{chaudhuri2010tla,tla_proofs} allows checking proofs manually written in TLA+, and has been used to verify several distributed protocols, including variants of Paxos~\cite{lamport2011byzantizing,chand2016formal}.

The derivation of inductive invariants for distributed protocols continues to be mostly carried out through refinement proofs using interactive theorem proving~\cite{chaudhuri2010verifying,hawblitzel2015ironfleet,Wilcox2015Verdi,padon2016ivy,druagoi2016psync,hoenicke2017thread,v2019pretend,kragl2020refinement}, which demands significant manual effort and profound domain expertise.
The first attempts at automatically deriving quantified invariants were reported in~\cite{pnueli2001automatic, arons2001parameterized}, using \emph{invisible invariants}.
The intuition underlying this method was the assumption that the system is ``sufficiently symmetric,'' and that its behavior can be captured by any $m$-subset of its processes as a universally-quantified invariant. However, universally-quantified invariants are not guaranteed to be inductive or to imply the safety property. Spatial regularity was further explored in~\cite{ip1993better, ip1996better,clarke1993exploiting,emerson1995reasoning,emerson1996symmetry,sistla2000smc} to reduce the verification of an $n$-process system to that of a \textit{quotient} system at a small \textit{cutoff} size.

Notwithstanding the undecidability result of Apt and Kozen~\cite{apt1986limits}, many efforts to automatically infer quantified inductive invariants have been reported with the pace increasing in recent years~\cite{karbyshev2017property,gurfinkel2018quantifiers,feldman2019inferring,ma2019i4,pldi20folic3,hance2021finding,yao2021distai}.
Verification of parameterized systems is further explored in~\cite{ghilardi2010backward,conchon2012cubicle,li2015paraverifier,abdulla2016parameterized,dooley2016proving}.
However, unlike \icpo, these methods generally do not scale to complex protocols like Lamport's Paxos, since these methods rely heavily on unbounded reasoning and are limited to specifications in the EPR fragment of first-order logic.

Our technique builds on these works, with the capability to automatically infer the required quantified inductive invariant using the latest advancements in model checking, by extending our recent work~\cite{goel2021on} on symmetry boosting and finite convergence with range boosting and hierarchical strengthening.

\section{Conclusions \& Future Work}
\label{sec:conclude}
\noindent We proposed \textit{range boosting}, a novel technique that extends the incremental induction algorithm to utilize the temporal regularity in distributed protocols through quantified reasoning over ordered ranges. 
We also presented \textit{hierarchical strengthening}, a simple technique that utilizes the hierarchical structure of protocol specifications to enable automatic verification of complex distributed protocols with high scalability.
Given the four-level hierarchy of the Paxos specification, we showed that these techniques, coupled with our recent work on symmetry boosting and finite convergence, provide, to our knowledge, the first demonstration of an automatically-inferred inductive invariant for the original Lamport's Paxos algorithm. 

While introducing \paxossimple and \paxosabstract to get the four-level Paxos hierarchy was quite easy, these intermediate levels were still added manually.
It is appealing to explore counterexample-guided abstraction-refinement (CEGAR) techniques~\cite{clarke2000counterexample,clarke2003counterexample}
to automatically identify these intermediate levels whenever needed to overcome complexity.
Specifically, investigating how to leverage clause learning feedback from incomplete runs to identify bottlenecks in proof inference and utilizing this information to automatically abstract away irrelevant details from the low-level protocol can help in making the complete procedure automatic end-to-end.
We leave this investigation as future work.

Exploring inference with existential quantifiers in range boosting can also be an interesting future direction, though intuitively, existential quantification over temporal behaviors looks unnecessary for proving safety properties.
Future work also includes automatically inferring inductive proofs for other distributed protocols, such as Byzantine Paxos~\cite{lamport2011byzantizing}, Raft~\cite{ongaro2014search}, etc., and exploring the verification of consensus algorithms in blockchain applications.

\section*{Data Availability Statement and Acknowledgments}
\addcontentsline{toc}{section}{Data Availability Statement and Acknowledgments}
\noindent The software and data sets generated and analyzed during the current study, including all experimental data, evaluation scripts, and \icpo source code are available at \url{https://github.com/aman-goel/fmcad2021exp}.

We thank Leslie Lamport for the TLA+ video course~\cite{tla_video_course}, which shaped several ideas presented in this paper. We thank the developers of TLA+~\cite{Kuppe_2019,tlagithub}, Yices~\cite{dutertre2014yices}, Z3~\cite{demoura2008z3}, pySMT~\cite{gario2015pysmt}, and Ivy~\cite{padon2016ivy} for making their tools openly available. 
We also thank the reviewers for their valuable comments.

\bibliographystyle{IEEEtran}
\bibliography{IEEEabrv,header-standard.bib,cite_database.bib,reference-db.bib}

\appendices

\clearpage
\onecolumn
\section*{Appendices}
\vspace{10pt}
\noindent We include additional/supplementary material in the appendices, as follows:
\vspace{10pt}
\begin{itemize}
    \item[] Appendix~\ref{app:sizes}: \textit{Finite instance sizes used in the experiments}
    \begin{itemize}
    \item[--] Lists down the instance sizes for \icpo and \ifour for each protocol in the evaluation (Section~\ref{sec:exp})
    \end{itemize}
    \vspace{10pt}
    
    \item[] Appendix~\ref{app:paxos_intermediate}: \textit{TLA+ description for \paxossimple and \paxosabstract}
    \begin{itemize}
    \item[--] Presents full TLA+ descriptions of \paxossimple and \paxosabstract
    \end{itemize}
    \vspace{10pt}
    
\end{itemize}

\vspace{30pt}
\section{Finite Instance Sizes used in the Experiments}
\label{app:sizes}

\vspace{20pt}
\subsection{Finite Instance Sizes for \icpo}
\label{app:ic3po_size}
\noindent Table~\ref{tab:finite_sizes_ic3po} lists down the initial base instance sizes used for \icpo runs in the evaluation (Section~\ref{sec:exp}) for each protocol. The table also includes the final $cutoff$ instance sizes reached, where the corresponding inductive invariant generalizes to be an inductive proof for \textit{any} size. Note again that \icpo checks for finite convergence and updates the instance sizes automatically, as detailed in~\cite{goel2021on}.

\vspace{10pt}
\begin{table}[H]
\setlength\tabcolsep{4pt}
\begin{center}
\begin{tabular}{ll|l}
	\hline
    \multicolumn{2}{c|}{Protocol} & \multicolumn{1}{c}{Finite instance sizes used for \icpo} \\
	\hline
	\multirow{3}{*}{\epr} & epr-paxos  & $\fin{value} = 2,~\fin{node} = 3,~\fin{quorum} = 3,~\fin{round} = 4$\\
	& epr-flexible\_paxos  & $\fin{value} = 2,~\fin{node} = 3,~\fin{quorum1} = 3,~\fin{quorum2} = 3,~\fin{round} = 4$\\
	& epr-multi\_paxos  & $\fin{value} = 2,~\fin{node} = 3,~\fin{quorum} = 3,~\fin{round} = 4,~\fin{inst} = 2,~\fin{votemap} = 2$\\
	\hline
	\multirow{6}{*}{\orig} & \voting  & $\fin{value} = 2,~\fin{acceptor} = 3,~\fin{quorum} = 3,~\fin{ballot} = 4$\\
	& \paxossimple  & $\fin{value} = 2,~\fin{acceptor} = 3,~\fin{quorum} = 3,~\fin{ballot} = 4$\\
	& \paxosabstract  & $\fin{value} = 2,~\fin{acceptor} = 3,~\fin{quorum} = 3,~\fin{ballot} = 4 \mapsto 5$\\
	& \paxos  & $\fin{value} = 2,~\fin{acceptor} = 3,~\fin{quorum} = 3,~\fin{ballot} = 4$\\
	& \multipaxos  & $\fin{value} = 2,~\fin{acceptor} = 3,~\fin{quorum} = 3,~\fin{ballot} = 4,~\fin{instances} = 2$\\
	& \flexiblepaxos  & $\fin{value} = 2,~\fin{acceptor} = 3,~\fin{quorum1} = 3,~\fin{quorum2} = 3,~\fin{ballot} = 4$\\
	\hline
\end{tabular}
\captionsetup{justification=centering, belowskip=0pt}
\caption{
Finite instance sizes used for \icpo
}
\captionsetup{justification=raggedright, aboveskip=0pt}
\caption*{
   $\s = x$ denotes sort $\s$ has both initial base size and final cutoff size $x$ \\ 
   $\s = x \mapsto y$ denotes sort $\s$ has initial size $x$ and final cutoff size $y$ (only happens for \paxosabstract)}
\label{tab:finite_sizes_ic3po}
\end{center}
\end{table}

\vspace{20pt}
\subsection{Finite Instance Sizes for \ifour}
\label{app:i4_size}
\noindent Table~\ref{tab:finite_sizes_i4} lists down the instance sizes used for \ifour runs in the evaluation (Section~\ref{sec:exp}) for each protocol.

\vspace{10pt}
\begin{table}[H]
\setlength\tabcolsep{4pt}
\begin{center}
\begin{tabular}{ll|l}
	\hline
    \multicolumn{2}{c|}{Protocol} & \multicolumn{1}{c}{Finite instance sizes used for \ifour} \\
	\hline
	\multirow{3}{*}{\epr} & epr-paxos  & $\fin{value} = 2,~\fin{node} = 3,~\fin{quorum} = 3,~\fin{round} = 4$\\
	& epr-flexible\_paxos  & $\fin{value} = 2,~\fin{node} = 3,~\fin{quorum1} = 3,~\fin{quorum2} = 3,~\fin{round} = 4$\\
	& epr-multi\_paxos  & $\fin{value} = 2,~\fin{node} = 3,~\fin{quorum} = 3,~\fin{round} = 4,~\fin{inst} = 2,~\fin{votemap} = 2$\\
	\hline
	\multirow{6}{*}{\orig} & \voting  & $\fin{value} = 2,~\fin{acceptor} = 3,~\fin{quorum} = 3,~\fin{ballot} = 4$\\
	& \paxossimple  & $\fin{value} = 2,~\fin{acceptor} = 3,~\fin{quorum} = 3,~\fin{ballot} = 4$\\
	& \paxosabstract  & $\fin{value} = 2,~\fin{acceptor} = 3,~\fin{quorum} = 3,~\fin{ballot} = 5$\\
	& \paxos  & $\fin{value} = 2,~\fin{acceptor} = 3,~\fin{quorum} = 3,~\fin{ballot} = 4$\\
	& \multipaxos  & $\fin{value} = 2,~\fin{acceptor} = 3,~\fin{quorum} = 3,~\fin{ballot} = 4,~\fin{instances} = 2$\\
	& \flexiblepaxos  & $\fin{value} = 2,~\fin{acceptor} = 3,~\fin{quorum1} = 3,~\fin{quorum2} = 3,~\fin{ballot} = 4$\\
	\hline
\end{tabular}
\captionsetup{justification=centering, belowskip=0pt}
\caption{
Finite instance sizes used for \ifour
}
\label{tab:finite_sizes_i4}
\end{center}
\end{table}

\clearpage
\section{TLA+ description for \paxossimple and \paxosabstract}
\label{app:paxos_intermediate}
This section presents the complete TLA+ description of \paxossimple and \paxosabstract.

\vspace{10pt}
\begin{multicols}{2}
\begin{algorithm}[H]
\batchmode 
\tlatex
\setstretch{0.93}
\tlasize
\setboolean{shading}{true}
\@x{\makebox[5pt][r]{\scriptsize \hspace{1em}}}\moduleLeftDash\@xx{ {\MODULE} \text{\paxossimple}}\moduleRightDash\@xx{\makebox[5pt][r]{\scriptsize \hspace{1em}}}%

\@pvspace{4.0pt}%
\@x{\makebox[10pt][r]{\scriptsize 1\hspace{0.8em}} {\CONSTANTS} \svalue , \sacceptor , \squorum}%

\@pvspace{4.0pt}%
\@x{\makebox[10pt][r]{\scriptsize 2\hspace{0.8em}} \sballot \.{\defeq} Nat
 \.{\cup} \{ -1 \}}%

\@pvspace{4.0pt}%
\@x{\makebox[10pt][r]{\scriptsize 3\hspace{0.8em}} {\VARIABLES} msg1a , msg1b , msg2a , msg2b , maxBal}%

\@pvspace{4.0pt}%
\@x{\makebox[10pt][r]{\scriptsize 4\hspace{0.8em}} msg1a \@s{12} \.{\in} \sballot \.{\rightarrow}
 {\BOOLEAN}}%

\@pvspace{4.0pt}%
\@x{\makebox[10pt][r]{\scriptsize \hspace{0.8em}} msg1b \@s{12} \.{\in} (\sacceptor \times \sballot) \.{\rightarrow}
 {\BOOLEAN}}%

\@pvspace{4.0pt}%
\@x{\makebox[10pt][r]{\scriptsize \hspace{0.8em}} msg2a \@s{12} \.{\in} (\sballot \times \svalue) \.{\rightarrow}
 {\BOOLEAN}}%

\@pvspace{4.0pt}%
\@x{\makebox[10pt][r]{\scriptsize \hspace{0.8em}} msg2b \@s{12} \.{\in} (\sacceptor \times \sballot \times \svalue) \.{\rightarrow}
 {\BOOLEAN}}%
 
\@pvspace{4.0pt}%
\@x{\makebox[10pt][r]{\scriptsize \hspace{0.8em}} maxBal\@s{7} \.{\in} \sacceptor \.{\rightarrow}
 \sballot}%

\@pvspace{4.0pt}%
\@x{\makebox[10pt][r]{\scriptsize 5\hspace{0.8em}} {\ASSUME}\@s{4} \.{\land}  \@s{4} \A\, Q \.{\in} \squorum \.{:} Q \subseteq \sacceptor } %
\@x{\makebox[10pt][r]{\scriptsize \hspace{0.8em}} \@s{34} \.{\land}  \@s{4} \A\, Q_1 , Q_2 \.{\in} \squorum \.{:} Q_1 \.{\cap} Q_2 \.{\neq} \{ \} } %

\@pvspace{4.0pt}%
\@x{\makebox[10pt][r]{\scriptsize 6\hspace{0.8em}} chosenAt(b, v) \.{\defeq} \E\, Q \.{\in} \squorum \.{:} \A\, A \.{\in} Q \.{:} msg2b(
 A , b , v )}
 
\@pvspace{4.0pt}%
\@x{\makebox[10pt][r]{\scriptsize 7\hspace{0.8em}} chosen(v) \.{\defeq} \E\, B \.{\in} \sballot \.{:} chosenAt(B , v )}
 
\@pvspace{4.0pt}%
\begin{shaded3}
\@x{\makebox[10pt][r]{\scriptsize 8\hspace{0.8em}} showsSafeAtSimplePaxos ( q , b , v ) \.{\defeq}}%
\@x{\makebox[10pt][r]{\scriptsize \hspace{0.8em}} \.{\land} \@s{1} \A\, A \.{\in} q \.{:} msg1b(A , b)}%
\@x{\makebox[10pt][r]{\scriptsize \hspace{0.8em}} \.{\land} \@s{2} \.{\lor} \@s{1} \A\, A \.{\in} \sacceptor \.{:} \A\, M_b \.{\in} \sballot \.{:} \A\, M_v \.{\in} \svalue \.{:} }
\@x{\@s{50} \neg (\, A \.{\in} q \@s{2} \.{\land} \@s{2} msg1b(A , b) \@s{2} \.{\land} \@s{2} msg2b(A, M_b, M_v) \,)}%
\@x{\@s{21} \.{\lor} \@s{1} \E\, M_b \.{\in} \sballot \.{:} } 
\@x{\@s{33} \.{\land} \@s{2} \E\, A \.{\in} q \.{:} msg1b(A , b) \@s{2} \.{\land} \@s{2} msg2b(A, M_b, v) }
\@x{\@s{33} \.{\land} \@s{2} \A\, A \.{\in} q \.{:} \A\, M_{b2} \.{\in} \sballot \.{:} \A\, M_{v2} \.{\in} \svalue \.{:} } 
\@x{\@s{48} msg1b(A , b) \@s{2} \.{\land} \@s{2} msg2b(A, M_{b2}, M_{v2}) \.{\rightarrow} M_{b2} \leq M_b}
\end{shaded3}

\@pvspace{4.0pt}%
\@x{\makebox[10pt][r]{\scriptsize 9\hspace{0.8em}} isSafeAtSimplePaxos(b , v) \.{\defeq} \E\, Q {\in} \squorum \.{:}}
\@x{\@s{125} showsSafeAtSimplePaxos(Q , b , v )}

\@pvspace{4.0pt}%
\@x{\makebox[10pt][r]{\scriptsize 10\hspace{0.8em}} Phase1a( b ) \@s{2} \.{\defeq}}
\@x{\makebox[10pt][r]{\scriptsize \hspace{0.8em}} \.{\land} \@s{2} b \.{\neq} -1}
\@x{\makebox[10pt][r]{\scriptsize \hspace{0.8em}} \.{\land} \@s{2} msg1a \.{'} \.{=} [ msg1a {\EXCEPT} {\bang} [ b ] \.{=} \top ]}%
\@x{\makebox[10pt][r]{\scriptsize \hspace{0.8em}} \.{\land} \@s{2} {\UNCHANGED}  msg1b , msg2a , msg2b , maxBal}%

\@pvspace{4.0pt}%
\@x{\makebox[10pt][r]{\scriptsize 11\hspace{0.8em}} Phase1b( a , b ) \@s{2} \.{\defeq}}
\@x{\makebox[10pt][r]{\scriptsize \hspace{0.8em}} \.{\land} \@s{2} b \.{\neq} -1 \@s{3} \.{\land} \@s{3} msg1a(b) \@s{3} \.{\land} \@s{3} b > maxBal(a)}
\@x{\makebox[10pt][r]{\scriptsize \hspace{0.8em}} \.{\land} \@s{2} maxBal \.{'} \.{=} [ maxBal {\EXCEPT} {\bang} [ a ] \.{=} b ]}%
\@x{\makebox[10pt][r]{\scriptsize \hspace{0.8em}} \.{\land} \@s{2} msg1b \.{'} \.{=} [ msg1b {\EXCEPT} {\bang} [a, b] \.{=} \top ]}%
\@x{\makebox[10pt][r]{\scriptsize \hspace{0.8em}} \.{\land} \@s{2} {\UNCHANGED} msg1a , msg2a , msg2b}%

\@pvspace{4.0pt}%
\@x{\makebox[10pt][r]{\scriptsize 12\hspace{0.8em}} Phase2a( b, v ) \@s{2} \.{\defeq}}
\@x{\makebox[10pt][r]{\scriptsize \hspace{0.8em}} \.{\land} \@s{2} b \.{\neq} -1 \@s{3} \.{\land} \@s{3} \neg(\, \E\, V \.{\in} \svalue \.{:} msg2a( b , V ) \,)}
\@x{\makebox[10pt][r]{\scriptsize \hspace{0.8em}} \.{\land} \@s{2} isSafeAtSimplePaxos(b, v)}
\@x{\makebox[10pt][r]{\scriptsize \hspace{0.8em}} \.{\land} \@s{2} msg2a \.{'} \.{=} [ msg2a {\EXCEPT} {\bang} [b, v] \.{=} \top ]}%
\@x{\makebox[10pt][r]{\scriptsize \hspace{0.8em}} \.{\land} \@s{2} {\UNCHANGED} msg1a , msg1b , msg2b , maxBal}%

\@pvspace{4.0pt}%
\@x{\makebox[10pt][r]{\scriptsize 13\hspace{0.8em}} Phase2b( a , b , v ) \@s{2} \.{\defeq}}
\@x{\makebox[10pt][r]{\scriptsize \hspace{0.8em}} \.{\land} \@s{2} b \.{\neq} -1 \@s{3} \.{\land} \@s{3} msg2a(b, v) \@s{3} \.{\land} \@s{3} b \.{\geq} maxBal(a)}
\@x{\makebox[10pt][r]{\scriptsize \hspace{0.8em}} \.{\land} \@s{2} maxBal \.{'} \@s{6} \.{=} [ maxBal {\EXCEPT} {\bang} [ a ] \.{=} b ]}%
\@x{\makebox[10pt][r]{\scriptsize \hspace{0.8em}} \.{\land} \@s{2} msg2b \.{'} \@s{12} \.{=} [ msg2b {\EXCEPT} {\bang} [a, b, v] \.{=} \top ]}%
\@x{\makebox[10pt][r]{\scriptsize \hspace{0.8em}} \.{\land} \@s{2} {\UNCHANGED} msg1a , msg1b , msg2a}%

\@pvspace{4.0pt}%
\@x{\makebox[10pt][r]{\scriptsize 14\hspace{0.8em}} \init \.{\defeq} \.\A\, A \.{\in} \sacceptor \.{:} B \.{\in} \sballot \.{:} }
\@x{\@s{43} \.{\land} \@s{2} {\neg} msg1a(B)}
\@x{\@s{43} \.{\land} \@s{2} {\neg} msg1b(A, B)}
\@x{\@s{43} \.{\land} \@s{2} \.\A\, V \.{\in} \svalue \.{:} {\neg} msg2a(B, V) \@s{3} \.{\land} \@s{3} {\neg} msg2b(A, B, V)}
\@x{\@s{43} \.{\land} \@s{2} maxBal(A) = -1} %

\@pvspace{4.0pt}%
\@x{\makebox[10pt][r]{\scriptsize 15\hspace{0.8em}} \nex \.{\defeq} \.\E\, A \.{\in} \sacceptor \.{:} B \.{\in} \sballot \.{:} V \.{\in} \svalue \.{:} }
\@x{\@s{43} \.{\lor} \@s{2} Phase1a(B) \@s{15} \.{\lor} \@s{3} Phase1b(A, B)}
\@x{\@s{43} \.{\lor} \@s{2} Phase2a(B, V) \@s{3} \.{\lor} \@s{3} Phase2b(A, B, V)}

\@pvspace{4.0pt}%
 \@x{\makebox[10pt][r]{\scriptsize 16\hspace{0.8em}} \resizebox{0.95\linewidth}{!}{$\prop \@s{0} \.{\defeq} \@s{0} \.\A\, V_1, V_2 {\in} \svalue \.{:} chosen(V_1) \.{\land} chosen(V_2) \.{\rightarrow} V_1 \.{=} V_2$}}%

 \@pvspace{4.0pt}%
\@x{\makebox[5pt][r]{\scriptsize \hspace{1em}}}\bottombar\@xx{\makebox[5pt][r]{\scriptsize \hspace{1em}}}%
\captionof{figure}{\paxossimple protocol in pretty-printed TLA+}
\label{fig:paxos_simple_full}
\end{algorithm}%

\begin{algorithm}[H]
\batchmode 
\tlatex
\setstretch{0.93}
\tlasize
\setboolean{shading}{true}
\@x{\makebox[5pt][r]{\scriptsize \hspace{1em}}}\moduleLeftDash\@xx{ {\MODULE} \text{\paxosabstract}}\moduleRightDash\@xx{\makebox[5pt][r]{\scriptsize \hspace{1em}}}%

\@pvspace{4.0pt}
\@x{\makebox[10pt][r]{\scriptsize 1\hspace{0.8em}} {\CONSTANTS} \svalue , \sacceptor , \squorum}%

\@pvspace{4.0pt}
\@x{\makebox[10pt][r]{\scriptsize 2\hspace{0.8em}} \sballot \.{\defeq} Nat
 \.{\cup} \{ -1 \}}%

\@pvspace{4.0pt}
\@x{\makebox[10pt][r]{\scriptsize 3\hspace{0.8em}} {\VARIABLES} msg1a , msg1b , msg2a , msg2b , maxBal}%

\@pvspace{4.0pt}
\@x{\makebox[10pt][r]{\scriptsize 4\hspace{0.8em}} msg1a \@s{12} \.{\in} \sballot \.{\rightarrow}
 {\BOOLEAN}}%

\@pvspace{4.0pt}
\@x{\makebox[10pt][r]{\scriptsize \hspace{0.8em}} msg1b \@s{12} \.{\in} (\sacceptor \times \sballot \times \sballot \times \svalue) \.{\rightarrow}
 {\BOOLEAN}}%

\@pvspace{4.0pt}
\@x{\makebox[10pt][r]{\scriptsize \hspace{0.8em}} msg2a \@s{12} \.{\in} (\sballot \times \svalue) \.{\rightarrow}
 {\BOOLEAN}}%

\@pvspace{4.0pt}
\@x{\makebox[10pt][r]{\scriptsize \hspace{0.8em}} msg2b \@s{12} \.{\in} (\sacceptor \times \sballot \times \svalue) \.{\rightarrow}
 {\BOOLEAN}}%
 
\@pvspace{4.0pt}
\@x{\makebox[10pt][r]{\scriptsize \hspace{0.8em}} maxBal\@s{7} \.{\in} \sacceptor \.{\rightarrow}
 \sballot}%

\@pvspace{4.0pt}
\@x{\makebox[10pt][r]{\scriptsize 5\hspace{0.8em}} {\ASSUME}\@s{4} \.{\land}  \@s{4} \A\, Q \.{\in} \squorum \.{:} Q \subseteq \sacceptor } %
\@x{\makebox[10pt][r]{\scriptsize \hspace{0.8em}} \@s{34} \.{\land}  \@s{4} \A\, Q_1 , Q_2 \.{\in} \squorum \.{:} Q_1 \.{\cap} Q_2 \.{\neq} \{ \} } %

\@pvspace{4.0pt}
\@x{\makebox[10pt][r]{\scriptsize 6\hspace{0.8em}} chosenAt(b, v) \.{\defeq} \E\, Q \.{\in} \squorum \.{:} \A\, A \.{\in} Q \.{:} msg2b(
 A , b , v )}
 
\@pvspace{4.0pt}
\@x{\makebox[10pt][r]{\scriptsize 7\hspace{0.8em}} chosen(v) \.{\defeq} \E\, B \.{\in} \sballot \.{:} chosenAt(B , v )}
 
\@pvspace{4.0pt}
\@x{\makebox[10pt][r]{\scriptsize 8\hspace{0.8em}} showsSafeAtPaxos ( q , b , v ) \.{\defeq}}%
\@x{\makebox[10pt][r]{\scriptsize \hspace{0.8em}} \.{\land} \@s{1} \A\, A \.{\in} q \.{:} \E\, M_b \.{\in} \sballot \.{:} \E\, M_v \.{\in} \svalue \.{:} msg1b(A , b , M_b , M_v)}%
\@x{\makebox[10pt][r]{\scriptsize \hspace{0.8em}} \.{\land} \@s{2} \.{\lor} \@s{1} \A\, A \.{\in} \sacceptor \.{:} \A\, M_b \.{\in} \sballot \.{:} \A\, M_v \.{\in} \svalue \.{:} }
\@x{\@s{50} \neg (\, A \.{\in} q \@s{2} \.{\land} \@s{2} msg1b(A , b , M_b , M_v) \@s{2} \.{\land} \@s{2} (M_b \.{\neq} -1) \,)}%
\@x{\@s{21} \.{\lor} \@s{1} \E\, M_b \.{\in} \sballot \.{:} } \@x{\@s{33} \.{\land} \@s{2} \E\, A \.{\in} q \.{:} msg1b(A , b , M_b , v) \@s{2} \.{\land} \@s{2} (M_b \.{\neq} -1) }
\@x{\@s{33} \.{\land} \@s{2} \A\, A \.{\in} q \.{:} \A\, M_{b2} \.{\in} \sballot \.{:} \A\, M_{v2} \.{\in} \svalue \.{:} } \@x{\@s{48} msg1b(A , b , M_{b2} , M_{v2}) \@s{2} \.{\land} \@s{2} (M_{b2} \.{\neq} -1) \.{\rightarrow} M_{b2} \leq M_b}

\@pvspace{4.0pt}
\@x{\makebox[10pt][r]{\scriptsize 9\hspace{0.8em}} isSafeAtPaxos(b , v) \.{\defeq} \E\, Q {\in} \squorum \.{:} showsSafeAtPaxos(Q , b , v )}

\@pvspace{4.0pt}
\@x{\makebox[10pt][r]{\scriptsize 10\hspace{0.8em}} Phase1a( b ) \@s{2} \.{\defeq}}
\@x{\makebox[10pt][r]{\scriptsize \hspace{0.8em}} \.{\land} \@s{2} b \.{\neq} -1}
\@x{\makebox[10pt][r]{\scriptsize \hspace{0.8em}} \.{\land} \@s{2} msg1a \.{'} \.{=} [ msg1a {\EXCEPT} {\bang} [ b ] \.{=} \top ]}%
\@x{\makebox[10pt][r]{\scriptsize \hspace{0.8em}} \.{\land} \@s{2} {\UNCHANGED}  msg1b , msg2a , msg2b , maxBal}%

\@pvspace{4.0pt}
\begin{shaded1}
\@x{\makebox[10pt][r]{\scriptsize  11\hspace{0.5em}} Phase1b( a , b ) \@s{2} \.{\defeq}}
\@x{\makebox[10pt][r]{\scriptsize \hspace{0.8em}} \.{\land} \@s{2} b \.{\neq} -1 \@s{3} \.{\land} \@s{3} msg1a(b) \@s{3} \.{\land} \@s{3} b > maxBal(a)}
\@x{\makebox[10pt][r]{\scriptsize \hspace{0.8em}} \.{\land} \@s{2} maxBal \.{'} \.{=} [ maxBal {\EXCEPT} {\bang} [ a ] \.{=} b ]}%
\@x{\makebox[10pt][r]{\scriptsize \hspace{0.8em}} \.{\land} \@s{2} \E\, M_b \.{\in} \sballot \.{:}  \E\, M_v \.{\in} \svalue \.{:} }
\@x{\@s{23} \.{\land} \@s{2} \.{\lor} \@s{2} \.{\land} \@s{2} (M_b = -1)}
\@x{\@s{48} \.{\land} \@s{2} \A\, B \.{\in} \sballot \.{:} \A\, V \.{\in} \svalue \.{:} \neg msg2b(a , B, V)}
\@x{\@s{36} \.{\lor} \@s{2} \.{\land} \@s{2} (M_b \neq -1) \@s{2} \.{\land} \@s{2} msg2b(a, M_b, M_v)}
\@x{\@s{48} \.{\land} \@s{2} \A\, B \.{\in} \sballot \.{:} \A\, V \.{\in} \svalue \.{:}}
\@x{\@s{108} msb2b(a, B, V) \.{\rightarrow} B \leq M_b}

\@x{\@s{23} \.{\land} \@s{2} msg1b \.{'} \.{=} [ msg1b {\EXCEPT} {\bang} [a, b, M_b, M_v] \.{=} \top ]}%
\@x{\makebox[10pt][r]{\scriptsize \hspace{0.8em}} \.{\land} \@s{2} {\UNCHANGED} msg1a , msg2a , msg2b}%
\end{shaded1}



\@pvspace{4.0pt}
\@x{\makebox[10pt][r]{\scriptsize 12\hspace{0.8em}} Phase2a( b, v ) \@s{2} \.{\defeq}}
\@x{\makebox[10pt][r]{\scriptsize \hspace{0.8em}} \.{\land} \@s{2} b \.{\neq} -1 \@s{3} \.{\land} \@s{3} \neg(\, \E\, V \.{\in} \svalue \.{:} msg2a( b , V ) \,)}
\@x{\makebox[10pt][r]{\scriptsize \hspace{0.8em}} \.{\land} \@s{2} isSafeAtPaxos(b, v)}
\@x{\makebox[10pt][r]{\scriptsize \hspace{0.8em}} \.{\land} \@s{2} msg2a \.{'} \.{=} [ msg2a {\EXCEPT} {\bang} [b, v] \.{=} \top ]}%
\@x{\makebox[10pt][r]{\scriptsize \hspace{0.8em}} \.{\land} \@s{2} {\UNCHANGED} msg1a , msg1b , msg2b , maxBal}%

\@pvspace{4.0pt}
\@x{\makebox[10pt][r]{\scriptsize 13\hspace{0.8em}} Phase2b( a , b , v ) \@s{2} \.{\defeq}}
\@x{\makebox[10pt][r]{\scriptsize \hspace{0.8em}} \.{\land} \@s{2} b \.{\neq} -1 \@s{3} \.{\land} \@s{3} msg2a(b, v) \@s{3} \.{\land} \@s{3} b \.{\geq} maxBal(a)}
\@x{\makebox[10pt][r]{\scriptsize \hspace{0.8em}} \.{\land} \@s{2} maxBal \.{'} \@s{6} \.{=} [ maxBal {\EXCEPT} {\bang} [ a ] \.{=} b ]}%
\@x{\makebox[10pt][r]{\scriptsize \hspace{0.8em}} \.{\land} \@s{2} msg2b \.{'} \@s{12} \.{=} [ msg2b {\EXCEPT} {\bang} [a, b, v] \.{=} \top ]}%
\@x{\makebox[10pt][r]{\scriptsize \hspace{0.8em}} \.{\land} \@s{2} {\UNCHANGED} msg1a , msg1b , msg2a}%

\@pvspace{4.0pt}
\@x{\makebox[10pt][r]{\scriptsize 14\hspace{0.8em}} \init \.{\defeq} \.\A\, A \.{\in} \sacceptor \.{:} B \.{\in} \sballot \.{:} }
\@x{\@s{43} \.{\land} \@s{2} {\neg} msg1a(B)}
\@x{\@s{43} \.{\land} \@s{2} \.\A\, M_b \.{\in} \sballot \.{:} M_v \.{\in} \svalue \.{:} {\neg} msg1b(A, B, M_b, M_v)}
\@x{\@s{43} \.{\land} \@s{2} \.\A\, V \.{\in} \svalue \.{:} {\neg} msg2a(B, V) \@s{3} \.{\land} \@s{3} {\neg} msg2b(A, B, V)}
\@x{\@s{43} \.{\land} \@s{2} maxBal(A) = -1} %

\@pvspace{4.0pt}
\@x{\makebox[10pt][r]{\scriptsize 15\hspace{0.8em}} \nex \.{\defeq} \.\E\, A \.{\in} \sacceptor \.{:} B \.{\in} \sballot \.{:} V \.{\in} \svalue \.{:} }
\@x{\@s{43} \.{\lor} \@s{2} Phase1a(B) \@s{15} \.{\lor} \@s{3} Phase1b(A, B)}
\@x{\@s{43} \.{\lor} \@s{2} Phase2a(B, V) \@s{3} \.{\lor} \@s{3} Phase2b(A, B, V)}

\@pvspace{4.0pt}%
 \@x{\makebox[10pt][r]{\scriptsize 16\hspace{0.8em}} \resizebox{0.95\linewidth}{!}{$\prop \@s{0} \.{\defeq} \@s{0} \.\A\, V_1, V_2 {\in} \svalue \.{:} chosen(V_1) \.{\land} chosen(V_2) \.{\rightarrow} V_1 \.{=} V_2$}}%

 \@pvspace{4.0pt}
\@x{\makebox[5pt][r]{\scriptsize \hspace{1em}}}\bottombar\@xx{\makebox[5pt][r]{\scriptsize \hspace{1em}}}%
\captionof{figure}{\paxosabstract protocol in pretty-printed TLA+}
\label{fig:paxos_abstract_full}
\end{algorithm}%

\end{multicols}


\end{document}